\documentclass[final,1p,times]{elsarticle}
\usepackage{bm}
\usepackage{epsfig}
\usepackage{amsfonts}
\usepackage{amsmath}
\usepackage{geometry}
\begin{document}
\def\ba{{\bf a}}
\def\bk{{\bf k}}
\def\bn{{\bf n}}
\def\bp{{\bf p}}
\def\bq{{\bf q}}
\def\br{{\bf r}}
\def\bv{{\bf v}}
\def\bx{{\bf x}}
\def\bz{{\bf z}}
\def\bM{{\bf M}}
\def\bP{{\bf P}}
\def\bR{{\bf R}}
\def\bK{{\bf K}}
\def\bJ{{\bf J}}
\def\bF{{\bf F}}
\def\e{\epsilon}
\def\calH{\mathcal H}
\def\calV{\mathcal V}
\def\calU{\mathcal U}
\def\calE{\mathcal E}
\def\la{\langle}
\def\ra{\rangle}
\def\pa{\partial}
\def\beq{\begin{equation}}
\def\eeq{\end{equation}}
\def\bdm{\begin{displaymath}}
\def\edm{\end{displaymath}}
\def\nn{\nonumber}
\begin{frontmatter}
\title{Dynamics of harmonically-confined systems: some rigorous
results}

 \author{Zhigang Wu\corref{cor1}}
 \author{Eugene Zaremba\corref{cor2}}
\address{Department of Physics, Queen's University,Kingston, ON, K7L 3N6, Canada}
 \cortext[cor1]{E-mail address: zwu@physics.queensu.ca}
 \cortext[cor2]{E-mail address: zaremba@sparky.phy.queensu.ca}

\begin{abstract}
In this paper we consider the dynamics of
harmonically-confined atomic gases. We present various general
results which are independent of particle statistics,
interatomic interactions and dimensionality. Of particular
interest is the response of the system to external perturbations
which can be either static or dynamic in nature. We prove an extended
Harmonic Potential Theorem which is useful in determining the
damping of the centre of mass motion when the 
system is prepared initially in a highly nonequilibrium state.
We also study the response of the gas to a dynamic external
potential whose position is made to
oscillate sinusoidally in a given direction.
We show in this case that either the energy absorption rate
or the centre of mass dynamics can serve as a probe of the optical
conductivity of the system.
\end{abstract}


                                      
\end{frontmatter}  
\tableofcontents
\section{\label{sec:intro}Introduction}
The atomic gases in many cold atom experiments are confined in 
harmonic traps. An important consequence of this kind of confinement is 
that, in the absence of any additional 
external perturbation, the centre of mass 
of the system oscillates about the centre of the trap in simple 
harmonic motion without dissipation. This particular collective 
oscillation is referred to as the centre of mass or dipole mode.
According to the generalized Kohn theorem~\cite{Brey,Li}, 
this behaviour 
is a generic property of a harmonically-confined system in 
which the interactions between particles depend only on their relative 
separation, and is independent of other intrinsic properties
such as dimensionality, quantum
statistics and the state of internal excitation. For these
reasons, the undamped dipole oscillation can in fact be used
to accurately determine the trapping
frequencies~\cite{Stamper-Kurn} in situations where the
experimental parameters defining the trapping potential are not
known precisely. An additional, but more subtle, implication 
of such confinement is the content of the so-called Harmonic 
Potential Theorem (HPT)~\cite{Dobson}. In essence, the HPT demonstrates 
the existence of a class of dynamical many-body states for which the 
probability density moves without change in shape. This theorem 
imposes important constraints on the form of approximate 
theories which deal with the dynamics of harmonically-confined 
many-body systems~\cite{Dobson,Zaremba,Griffin}. 

When the harmonicity of the confining potential is 
compromised, however, the centre of mass is coupled to the 
internal degrees of freedom and its dynamics becomes sensitive to 
the intrinsic properties of the system, including the specific
form of the particle interactions. For this reason, the dipole 
oscillation can serve as an experimental diagnostic of various 
perturbations acting on the system. For instance, several 
experiments~\cite{Lye, Dries, Chen} have used dipole oscillations to 
study the transport of a Bose-condensate through a disordered medium or 
past a localized impurity. Although the motion of the condensate in these 
experiments does not lose its collectivity, dissipation does occur and 
leads to the damping of the centre of mass motion. Another experimental 
example is provided by the dipole oscillation of a trapped Bose gas in 
the presence of an optical lattice
potential~\cite{Berger01,Cataliotti03,Fertig,Rigol}. Here 
it was found the that dimensionality of the Bose gas plays a critical 
role in determining the way in which the centre of mass behaves
as a function of time.

In all of these experiments, the dipole oscillation of the atomic 
system is initiated by an abrupt displacement of the trapping 
potential along a certain axial direction. If the displacement 
is large, the system begins its evolution in a highly non-equilibrium
initial state. It is partly for this reason that much of the 
theoretical work dealing with the collective dynamics of Bose-condensed 
systems relies on numerical simulations of the time-dependent 
Gross-Pitaevskii (GP) equation~\cite{Modugno,Albert1,Albert2}. 
One of the goals of this paper is to show that this
nonequilibrium dynamics in the presence of the external
perturbation can be considered from a different point of view 
when the system is harmonically confined. By means of an 
appropriate transformation, one can equivalently think
of the system as being driven out of an initial {\it equilibrium}
state by a {\it dynamic} external perturbation oscillating
sinusoidally at the frequency of the trap. The availability of
this alternate point of view is a consequence of what we refer
to as the {\it extended} HPT. Its advantage
is that the external perturbation can be
treated by conventional linear response theory, at least when 
the perturbation is sufficiently weak. This approach
was used effectively in an earlier paper~\cite{Wu} to determine the 
damping of the centre of mass motion due to a disorder potential.

A second purpose of this paper is to study the response of a 
harmonically-confined system to an external potential which is
made to oscillate at an 
{\it arbitrary} frequency. Our discussion is motivated by a recent 
proposal~\cite{Tokuno} to probe the optical conductivity of a cold 
atomic gas in an optical lattice by shaking the lattice periodically 
along a certain direction. This is an interesting idea since it
provides a method of addressing experimentally the optical 
conductivity of a 
system consisting of {\it neutral} atoms. However, the authors of 
Ref.~\cite{Tokuno} only considered bosons within a {\it uniform} 
lattice, while in most experiments the atoms are also subjected
to a harmonic potential. In this paper, we show that one can also 
probe directly the optical conductivity of a gas that experiences 
a {\it combination} of a harmonic trapping potential and an 
arbitrary external potential when the latter is made to oscillate 
sinusoidally with a small amplitude. This generalization provides 
a precise link between theoretical calculations of the 
optical conductivity and possible experimental measurements on
harmonically-confined gases in the presence of various external
perturbations. 
 
The rest of the paper is organized as follows. In Sec.~2, we 
provide a basic discussion of the dipole modes of a 
harmonically-confined system. In Sec.~3, we consider the response 
of a harmonically-confined system to a time-dependent homogeneous
force. An explicit expression for the evolution operator of the
system is obtained which motivates the introduction of a rather useful 
unitary displacement operator. These results are then used to
provide an alternative derivation of the HPT. 
We next consider in Sec.~4 situations in which the system is
perturbed by an additional external potential that couples the
centre of mass and internal degrees of freedom. Here we present
a derivation of the extended HPT. In Sec.~5, we consider the 
energy absorption rate and centre of mass dynamics of a 
harmonically-confined gas that is subjected to an oscillating external 
potential and demonstrate that both aspects serve to probe 
the optical conductivity of the system. All of our findings
are summarized in Sec.~6.  

\section{\label{sec:HPT}Harmonically confined systems and dipole modes}
As a preliminary to the the development of the HPT and its extension, 
we discuss in this section the dipole modes of a harmonically-confined 
system and the underlying physics for the existence of such modes. The 
dipole modes are the low-lying collective excitations that have 
frequencies equal to the frequencies of the trap. Unlike other 
low-lying excitations, the frequencies of these modes are independent 
of the total number of trapped atoms and the atomic interactions. For
Bose-condensed systems, these modes are often discussed in the context 
of mean-field theory, namely as solutions to the time-dependent GP 
equation. However, it can be shown 
rigorously that such dipole modes exist for any harmonically-confined 
(bosonic or fermionic) system in which interactions depend only on the 
relative coordinates of the particles. The fundamental reason behind 
this is that the centre of mass degree of freedom is separable from all 
the internal degrees of freedom, which implies that there are 
excitations associated solely with motion of the centre of mass.

To demonstrate this, we consider a harmonically-confined many-body 
system described by the generic Hamiltonian
\beq
\hat H_0 =\sum_{i=1}^N \left (\frac{\hat \bp_i^2}{2m}+V_{\rm trap}(\hat 
\br_i)\right )+\sum_{i<j}v(\hat \br_i-\hat \br_j),
\label{H5}
\eeq
where the harmonic trapping potential is
\beq
V_{\rm trap}(\br) =
\frac{1}{2}m\sum_{\mu=x,y,z}\omega_\mu^2 r_\mu^2.
\eeq
We define the centre of mass co-ordinate $\hat 
{\bf R} = \frac{1} {N} \sum_{i=1}^N \hat \br_i$ and the total
momentum operator of the system $\hat \bP=\sum_i \hat\bp_i$. 
Since these two operators satisfy the commutation relation $[\hat 
R_\mu, \hat P_\nu]=i\hbar \delta_{\mu\nu}$, they are canonically
conjugate variables. Introducing the relative
variables $\hat\br'_i=\hat\br_i-\hat\bR$ and $\hat \bp'_i=\hat
\bp_i-\hat \bP/N$~\footnote{It should be noted that these
relative variables are {\it not} independent and are therefore
not canonically conjugate.}, we observe that the 
Hamiltonian $\hat H_0$ can be written as 
\beq
\hat H_0 =\hat H_{\rm cm}+\hat H_{\rm int}.
\eeq
Here
\beq
\hat H_{\rm cm}=\frac{\hat 
\bP^2}{2M}+\frac{1}{2}M\sum_{\mu=x,y,z}\omega^2_\mu \hat R_\mu^2
\label{H_cm}
\eeq
and
\beq
\hat H_{\rm int}=\sum_{i=1}^N \left (\frac{\hat \bp'_i}{2m}+V_{\rm 
trap}(\hat \br'_i)\right )+\sum_{i<j}v(\hat \br'_i-\hat \br'_j),
\eeq
where $M=Nm$ is the total mass of the system. 
The Hamiltonian for the centre of mass degree of freedom 
$\hat H_{\rm cm}$ is that of a harmonic oscillator; 
$\hat H_{\rm int}$ is the Hamiltonian determining 
the internal dynamics of the system. One can check that the centre 
of mass and relative variables commute, namely
\beq
[\hat R_\mu, \hat p_{i\nu}']=0;\quad [\hat P_\mu, \hat r_{i\nu}']=0.
\label{RP_com}
\eeq
It follows from these results that the centre of mass
Hamiltonian $\hat H_{\rm cm}$ commutes with $\hat H_{\rm int}$.
This means that the motion of the centre of mass decouples from
the internal dynamics of the system.

One simple implication of this decoupling is that the centre of 
mass exhibits simple harmonic motion at the frequencies 
of the trap. Using Eq.~(\ref{H_cm}) and (\ref{RP_com}), 
the Heisenberg equations of motion for the centre of mass 
coordinate and the total momentum are
\begin{align}
\label{eqR}
  \frac{d\hat{R}_{\mu,\rm I}(t)}{dt}&=\frac{1}{i\hbar}[\hat{R}_{\mu,\rm 
I}(t), \hat {H_0}]=\frac {\hat {P}_{\mu,\rm I}(t)}{M}, \\
 \frac{d\hat{P}_{ \mu,\rm I}(t)}{dt}&=\frac{1}{i\hbar}[\hat{P}_{\mu,\rm 
I}(t),\hat{H_0}]=-M\omega_\mu^2\hat{R}_{\mu,\rm I}(t),
 \label{eqP}
\end{align}
where $\hat R_{\mu,\rm I} (t)\equiv e^{i\hat H_0t/\hbar}\hat R_\mu 
e^{-i\hat H_0 t/\hbar}$ and $\hat P_{\mu,\rm I} (t)\equiv e^{i\hat 
H_0t/\hbar}\hat P_\mu e^{-i\hat H_0 t/\hbar}$. Equations (\ref{eqR}) 
and (\ref{eqP}) lead to the simple harmonic motion equation
\beq
\frac{d^2 \hat { R}_{\mu,\rm I}(t)}{d t^2}+\omega_\mu^2 \hat R_{\mu,\rm 
I}(t)=0.
\eeq
The formal solution of this equation is
\begin{align}
\label{solR}
 \hat {R}_{\mu,\rm I}(t)&=\hat {R}_\mu \cos\omega_\mu t+\frac{\hat 
{P}_\mu}{M\omega_\mu}\sin\omega_\mu t, \\
 \hat {P}_{\mu,\rm I}(t)&=-M\omega_\mu\hat {R}_\mu \sin\omega_\mu 
t+\hat {P}_\mu\cos\omega_\mu t.
 \label{solP}
\end{align}
From this we see that the expectation value $R_{\mu}(t)=\la\Psi(t)|\hat 
R_\mu|\Psi(t)\ra$ for an arbitrary state $|\Psi(t)\ra$ evolves in time 
according to the equation
\beq
R_\mu(t)=R_\mu(0)\cos\omega_\mu t +\frac{P_\mu 
(0)}{M\omega_\mu}\sin\omega_\mu t=A\cos (\omega_\mu t+\phi_0),
\label{cm_osc}
\eeq
where the amplitude $A$ and phase angle $\phi_0$ are determined by the 
initial conditions $R_\mu(0)$ and $P_\mu(0)$. This undamped harmonic 
oscillation of the centre of mass coordinate is the dipole oscillation 
we have been referring to.

The centre of mass excitations are conveniently described by
defining the centre of mass annihilation and creation operators
\beq
\hat a_{\mu}=\sqrt{\frac{M\omega_\mu}{2\hbar}}\left (\hat  R_\mu 
+\frac{i}{M\omega_\mu }\hat P_\mu\right ); \quad \hat 
a^\dag_{\mu}=\sqrt{\frac{M\omega_\mu}{2\hbar}}\left (\hat  R_\mu 
-\frac{i}{M\omega_\mu }\hat P_\mu\right ).
\label{a_cm}
\eeq
In terms of these operators, the Hamiltonian $\hat H_0$ takes
the form
\beq
\hat H_0=\sum_\mu \hbar \omega_\mu\left (\hat 
a^\dag_{\mu}\hat a_\mu + \frac{1}{2}\right ) +\hat H_{\rm int}.
\eeq
From Eq.~(\ref{RP_com}) and the definition in Eq.~(\ref{a_cm}), we 
observe that $\hat a_\mu$ and $\hat a^\dag_\mu$ commute with $\hat 
H_{\rm int}$.
Taking $|\Psi_\alpha\ra$ to be an eigenstate state of $\hat H_0$ with 
energy $E_\alpha$, we find that the state 
$|\Psi\ra=\frac{1}{\sqrt{n!}}\left (a^\dag_{\mu}\right 
)^n|\Psi_\alpha\ra$ satisfies
\beq
\hat H_0 |\Psi\ra=(E_\alpha+n\hbar\omega_\mu)|\Psi\ra,
\eeq
that is, $|\Psi\ra$ remains an eigenstate of $\hat H_0$ with energy 
$E=E_\alpha +n\hbar\omega_\mu$; the application of $\hat a^\dag_\mu$ 
creates a quantum of excitation of the centre of mass oscillation with 
energy $\hbar\omega_\mu$.

Finally, we mention the well known fact~\cite{Sakurai} that a simple 
harmonic oscillator has wave packet quantum states that move 
harmonically without any change in shape. These states are the 
so-called coherent states. Since the centre of mass degree of freedom 
is effectively a harmonic oscillator, analogous states also exist for 
the many-body system described by the Hamiltonian $\hat H_0$ in 
Eq.~(\ref{H5}). This property is encapsulated by the HPT discussed in 
the next section.

\section{Centre of mass motion in the presence
of an external driving force: the Harmonic Potential Theorem}
The separability of the centre of mass and internal degrees of
freedom leads to excitations which are associated solely with the
motion of the centre of mass. We now discuss a further implication of 
this property, namely the existence of a class of dynamical many-body 
states for which the probability density moves without change in shape. 
As shown by Dobson in his proof of the Harmonic Potential
Theorem~\cite{Dobson}, this behaviour can occur not only for
systems described by the Hamiltonian in Eq.~(\ref{H5}), but also 
when the system is subjected to an arbitrary time-dependent, but 
spatially homogeneous, force. In this situation, the system is
described by the Hamiltonian 
\beq
\hat H_F(t)=\hat H_0-\bF(t)\cdot\sum_{i=1}^N\hat \br_i=\hat 
H_0-N\bF(t)\cdot\hat \bR.
\label{Hwf}
\eeq
In the presence of the force ${\bf F}(t)$,
the system is still subjected at any instant of time to a purely
harmonic, albeit time-dependent, confining potential. It is in
this sense that we use the phrase ``purely harmonic confinement" to
distinguish this situation from those we consider later in which
additional perturbing potentials are also present.
Before dealing with these situations, we first determine the
dynamical state $|\Psi(t)\ra$ which evolves in time according to
the Hamiltonian $\hat H_F(t)$. This result will then be used to
provide an alternative derivation of the HPT.

The time evolution of $|\Psi(t)\ra$ is 
formally given by 
\beq
|\Psi(t)\ra=\hat \calU(t)|\Psi(0)\ra,
\label{dystate}
\eeq
where the unitary evolution operator $\hat \calU(t)$ satisfies the 
equation
\beq
i\hbar \frac{\pa}{\pa t}\hat \calU(t)=\hat H_F(t)\hat \calU(t)
\eeq
with the initial condition $\hat \calU(0)=\hat I$. To determine  $\hat 
\calU(t)$, we go to the interaction picture and define 
\beq
\hat \calU_{\rm I}(t)=e^{i\hat H_0 t/\hbar}\hat \calU(t),
\label{evoopint}
\eeq
which also has the initial condition $\hat \calU_{\rm I}(0)=\hat I$. 
This evolution operator satisfies the equation
\beq
i\hbar \frac{\pa}{\pa t}\hat \calU_{\rm I}(t)=-N\sum_\mu F_\mu (t)\hat 
R_{\mu,\rm I} (t) \hat \calU_{\rm I}(t),
\label{evolopeq}
\eeq
where $\hat R_{\mu,\rm I} (t)$  is given explicitly in 
Eq.~(\ref{solR}). The formal solution of Eq.~(\ref{evolopeq}) can be 
written as
\begin{align}
\hat \calU_{\rm I}(t)&=\lim_{N_s\rightarrow
\infty}\prod_{j=0}^{N_s-1} e^{ 
\frac{i}{\hbar}N\sum_\mu F_\mu(j\Delta t)\hat R_{\mu,\rm I}(j\Delta 
t)\Delta t} \nn \\
&=\lim_{N_s\rightarrow \infty}\prod_{\mu}e^{ 
\frac{i}{\hbar}NF_\mu((N_s-1)\Delta t)\hat R_{\mu,\rm I}((N_s-1)\Delta t)\Delta 
t} \cdots e^{ \frac{i}{\hbar}NF_\mu (\Delta t)\hat R_{\mu,\rm 
I}(\Delta t)\Delta t} e^{ \frac{i}{\hbar}NF_\mu(0)\hat 
R_{\mu,\rm I}(0)\Delta t},
\label{UIfsol}
\end{align}
where $\Delta t=t/N_s$. The second line follows from the fact that 
$\hat R_{\mu,\rm I} (t)$ and $\hat R_{\nu,\rm I} (t')$ commute
when $\mu\neq \nu$.

The product of operators in Eq.~(\ref{UIfsol}) can be evaluated 
recursively. Starting from the right and making use of the 
Baker-Hausdorff formula~\footnote{The Baker-Hausdorff formula 
states that $e^{\hat A+\hat B}=e^{\hat A}e^{\hat B}e^{-\hat C/2}$ if 
$\hat A$ and $\hat B$ commute with $\hat C=[\hat A,\hat B]$.}, we find
\begin{align}
&\exp\left\{ \frac{i}{\hbar}NF_\mu(\Delta t)\hat R_{\mu,\rm I}(\Delta 
t)\Delta t\right \} \exp \left \{ \frac{i}{\hbar}NF_\mu(0)\hat 
R_{\mu,\rm I}(0)\Delta t\right \} \nn \\
=&\exp \left \{ \frac{i}{\hbar}N\left[F_\mu(\Delta t)\hat R_{\mu,\rm 
I}(\Delta t)+F_\mu(0)\hat R_{\mu,\rm I}(0)\right ]\Delta 
t\right \}\exp\left \{\frac{iN}{2m\hbar\omega_\mu}F_\mu(\Delta t)F_\mu 
(0)\sin\omega_\mu\Delta t(\Delta t)^2\right \}.
\end{align}
To arrive at this result we have noted that
\begin{align}
[\hat R_{\mu,\rm I}(t), \hat R_{\mu,\rm 
I}(t')]=\frac{i\hbar}{Nm\omega_\mu}\sin\omega_\mu(t'-t),
\label{R_commu}
\end{align}
which is obtained using Eq.~(\ref{solR}).
After repeating these steps $j-1$ times, we must consider in the next 
step
\begin{align}
&\exp\left\{ \frac{i}{\hbar}NF_\mu(j\Delta t)\hat R_{\mu,\rm I}(j\Delta 
t)\Delta t\right \} \exp \left \{ 
\frac{i}{\hbar}N\sum_{k=0}^{j-1}F_\mu(k\Delta t)\hat R_{\mu,\rm 
I}(k\Delta t)\Delta t\right \} \nn \\
=&\exp \left \{ \frac{i}{\hbar}N\sum_{k=0}^jF_\mu(k\Delta t)\hat 
R_{\mu,\rm I}(k\Delta t)\Delta t \right \}\exp\left 
\{\frac{iN}{2m\hbar\omega_\mu}F_\mu(j\Delta t)\sum_{k=0}^{j-1}F_\mu 
(k\Delta t)\sin\omega_\mu(j\Delta t-k\Delta t)(\Delta t)^2\right \}.
\end{align}
It is clear that a phase factor of the kind given by the second 
exponential appears at each step of the process. 
Accumulating these phase factors and converting the summation into an 
integral, we find that Eq.~(\ref{UIfsol}) becomes
\begin{align}
\hat\calU_{\rm I}(t)
&=\exp\left\{\frac{i}{\hbar}\int_0^t dt' N\bF(t')\cdot\hat 
\bR_{\rm I}(t')\right \}  \nn \\
&\quad\times\prod_\mu \exp\left \{
\frac{iN}{2m\hbar\omega_\mu}\int_0^t dt'\int_{0}^{t'}dt''
F_\mu (t')F_\mu(t'')\sin\omega_\mu ( t'-t'')\right \}.
\label{U_I}
\end{align}
This result for $\hat\calU_{\rm I}(t)$ defines the 
dynamic state in Eq.~(\ref{dystate}) for an arbitrary initial
state $|\Psi(0)\ra$. 

With $\hat \bR_{\rm I}(t)$ given by Eq.~(\ref{solR}), the first
exponential factor in Eq.~(\ref{U_I}) takes the form of 
the unitary displacement operator
\beq
\hat{T}(\bx,\bp)\equiv \exp\left \{\frac{i}{\hbar}\left (\bp\cdot \hat 
{\bR}-\bx\cdot\hat {\bP}\right )\right \},
\label{T_ope}
\eeq
where $\bx$ is a position vector and $\bp$ is a momentum vector. 
This operator can be viewed as a generalization of the usual 
translation operator $\exp\{-i\bx\cdot\hat \bP
/\hbar\}$~\cite{Sakurai}. When
applied to some arbitrary state, the operator $\hat T(\bx,\bp)$ 
has the effect of shifting the state by $\bx$ in 
position space and by $\bp/N$ in momentum space. To see this, we make 
use of the Baker-Hausdorff formula to obtain
 \begin{align}
 \hat{T}(\bx,\bp)=\exp\left\{-\frac{i}{2\hbar}\bx\cdot 
\bp\right\}\exp\left \{\frac{i}{\hbar}\bp\cdot \hat {\bR}\right 
\}\exp\left\{-\frac{i}{\hbar}\bx\cdot\hat {\bP}\right \}.
 \label{Txpbh1}
 \end{align} 
 Defining the state $|\Psi'\ra=\hat T(\bx,\bp)|\Psi\ra$ and using 
Eq.~(\ref{Txpbh1}), one finds 
 \begin{align}
\Psi'(\br_1,\cdots,\br_N)&=\la \br_1,\cdots,\br_N|\Psi'\ra\nn\\
&=\exp\left[\frac{i}{\hbar}\bp\cdot(\bR-\bx/2)
\right ]\Psi(\br_1-\bx,\cdots,\br_N-\bx),
 \label{psi'r}
 \end{align}
 where $\bR=\sum_{i=1}^N \br_i /N$. We thus have 
$|\Psi'(\br_1,...,\br_N)|^2 =|\Psi(\br_1-\bx,...,\br_N-\bx)|^2$. 
Similarly, by interchanging the final two exponentials in
Eq.~(\ref{Txpbh1}), one can show that the momentum-space
wavefunction is
 \begin{align}
 \tilde \Psi'(\bp_1,\cdots,\bp_N)&=\la \bp_1,\cdots,\bp_N|\Psi'\ra \nn 
\\
 &=\exp\left[-\frac{i}{\hbar}\bx\cdot(\bP-\bp/2)\right 
]\tilde\Psi(\bp_1-\bp/N,\cdots,\bp_N-\bp/N),
 \label{psi'p}
 \end{align}
  where $\bP=\sum_{i=1}^N \bp_i$. Thus $|\tilde 
\Psi'(\bp_1,...,\bp_N)|^2 = 
|\tilde\Psi(\bp_1-\bp/N,...,\bp_N-\bp/N)|^2$, which implies that the 
{\it total} momentum of the state is boosted by $\bp$.
Furthermore, it is straightforward to demonstrate the operator
displacement properties
 \begin{align}
 \label{TfT}
 \hat{T}^\dag(\bx,\bp)\sum_{i=1}^Nf(\hat 
\br_i)\hat{T}(\bx,\bp)&=\sum_{i=1}^Nf(\hat \br_i+\bx), \\ 
 \hat{T}^\dag(\bx,\bp)\sum_{i=1}^Nf(\hat 
\bp_i)\hat{T}(\bx,\bp)&=\sum_{i=1}^Nf(\hat \bp_i+\bp/N).
 \label{TfT2}
 \end{align}
We will make use of these transformation properties in the following.

We now consider the dynamical evolution of the system when prepared
in the initial state
\beq
|\Psi(0)\ra=\hat{T}(\bx,\bp)|\Phi\ra,
\label{Psi_0_a}
\eeq 
where $|\Phi\ra$ is an arbitrary many-body state.
As discussed above,
this initial state is simply the state 
$|\Phi\ra$ translated rigidly in position space through 
the vector $\bx$ and given a total momentum boost of $\bp$.
At the end of this section we shall explain how such an initial 
state can be realized in cold atom experiments. 

We next show how Eq.~(\ref{U_I}) together with
Eq.~(\ref{Psi_0_a}) leads to the HPT. We 
have
\begin{align}
|\Psi(t)\ra&=e^{-i \hat H_0 t/\hbar}\hat\calU_{\rm I}(t)|\Psi(0)\ra, 
\nn \\
&=\left \{ e^{-i \hat H_0 t/\hbar}\calU_{\rm I}(t)\hat 
T(\bx,\bp)e^{i\hat H_0 t/\hbar}\right \}e^{-i\hat H_0
t/\hbar}|\Phi\ra.
\label{dystateII}
\end{align}
Here, the product $\hat\calU_{\rm I}(t)\hat T(\bx,\bp)$ can again
be evaluated using the Baker-Hausdorff formula. We find
 
\beq
\calU_{\rm I}(t)\hat T(\bx,\bp)=e^{-iS(t)/\hbar} \exp 
\left\{\frac{i}{\hbar}\left [ \bp\cdot\hat \bR-\bx\cdot\hat\bP+\int_0^t 
dt' N\bF(t')\cdot\hat\bR_{\rm I}(t')\right ] \right \},
\label{UT}
\eeq
where the phase $S(t)$ is
\beq
S(t)=-\chi(t) -\sum_\mu\frac{N}{2m\omega_\mu}\int_0^t 
dt'\int_{0}^{t'}dt'' F_\mu(t')F_\mu(t'')\sin\omega_\mu (
t'-t''),
\eeq
with
\beq
\chi(t) = \frac{1}{2}\sum_\mu \int_0^t dt' NF_\mu(t')\left [ x_\mu \cos
\omega_\mu t' +\frac{p_\mu}{M\omega_\mu}\sin \omega_\mu t'
\right ]
\eeq
Substituting Eq.~(\ref{UT}) into Eq.~(\ref{dystateII}), we find 
\begin{align}
|\Psi(t)\ra&=e^{-iS(t)/\hbar}\exp 
\left\{\frac{i}{\hbar}\left [ \bp\cdot\hat \bR_{\rm 
I}(-t)-\bx\cdot\hat\bP_{\rm I}(-t)+\int_0^t dt' 
N\bF(t')\cdot\hat\bR_{\rm I}(t'-t)\right ] \right
\}e^{-i\hat H_0 t/\hbar}|\Phi\ra \nn 
\\
&=e^{-iS(t)/\hbar}\hat T\left (\bx(t),\bp(t)\right )
e^{-i\hat H_0 t/\hbar}|\Phi\ra,
\label{dynamicstate}
\end{align}
where Eqs.~(\ref{solR}) and (\ref{solP}) are used to obtain the final 
result.
Here $x_\mu(t)$ and $p_\mu(t)$ are given by
\begin{align}
\label{xt}
x_\mu(t)&=x_{0\mu}(t)+\frac{N}{M\omega_\mu}\int_0^t dt' 
\sin\omega_\mu(t-t')F_\mu(t'),  \\
p_\mu(t)&=p_{0\mu}(t)+N\int_0^t dt' \cos \omega_\mu (t-t')F_\mu(t'),
\label{pt}
\end{align}
where 
\begin{align}
\label{x0t}
x_{0\mu}(t)& = x_\mu \cos\omega_\mu t 
+\frac{p_\mu}{M\omega_\mu}\sin\omega_\mu t,  \\
p_{0\mu}(t)& = - M\omega_\mu x_\mu\sin\omega_\mu t+p_\mu\cos\omega_\mu 
t.
\label{p0t}
\end{align}
From these expressions we see that $x_\mu(t)$ is simply the solution of 
the forced harmonic oscillator equation 
\beq
M\ddot x_\mu(t)+M\omega_\mu^2 x_\mu(t)=NF_\mu(t)
\label{cdoeq}
\eeq
with the initial conditions $x_\mu(0)=x_\mu$ and 
$\dot{x}_\mu(0)=p_\mu/M$. Likewise, 
$x_{0\mu}(t)$ is the solution in the absence of the force, again
with the initial conditions $x_{0\mu}(0)=x_\mu$ and 
$\dot{x}_{0\mu}(0)=p_\mu/M$. In terms of the solution of 
Eq.~(\ref{cdoeq}), the phase $S(t)$ can be simplified as
\beq
S(t)=-\frac{N}{2}\int_0^t dt' \bF(t')\cdot\bx(t').
\eeq
Equivalently, this can be written as 
\beq
S(t)=\int_0^t dt' \sum_{\mu}\left [\frac{1}{2}M\dot 
x_\mu^2(t')-\frac{1}{2}M\omega_\mu^2x_\mu^2(t') \right ],
\eeq
which is the classical action of a harmonic oscillator.

Equation (\ref{dynamicstate}) shows that the evolution of the
initial state in Eq.~(\ref{Psi_0_a}) can be considered as taking
place in two steps. First, the state $|\Phi\ra$ evolves {\it 
freely} for a time $t$, that is, in the absence of ${\bf F}(t)$.
The state $|\Phi(t)\ra \equiv e^{-i\hat H_0 t/\hbar}|\Phi\ra$ 
is then displaced by $\hat T\left
(\bx(t),\bp(t)\right )$ to generate, apart from a phase factor,
the final dynamical state of interest. Since the 
operator $\hat{T}(\bx(t),\bp(t))$ shifts a state in position space by 
$\bx(t)$, the wave function corresponding to the state $|\Psi (t)\ra$ 
in Eq.~(\ref{dynamicstate}) is 
$\Psi(\br_1,\cdots,\br_N;t)=e^{i\theta(t)}\Phi(\br_1-\bx(t),
\cdots,\br_N-\bx(t);t)$, where $\theta(t)$ is some time-dependent 
phase angle. Dobson's HPT now follows by choosing $|\Phi\ra$ to
be $|\Psi_\alpha\ra$, an eigenstate of $\hat H_0$. In this case
one has
$\Psi(\br_1,\cdots,\br_N;t)=e^{i\bar\theta(t)}\Psi_\alpha(\br_1-\bx(t),
\cdots,\br_N-\bx(t))$, which implies that
the probability density simply moves rigidly, following 
the trajectory of the centre of mass motion given by $\bx(t)$ in 
Eq.~(\ref{xt}). This property does not apply to an arbitrary 
initial state which evolves according to Eq.~(\ref{dystate}).

So far we have only discussed the evolution of a many-body
eigenstate. More 
generally, we can consider a system described by a statistical density 
matrix of the form
\beq
\hat \rho_0=\sum_k\lambda_k|\xi_k\ra\la\xi_k|.
\eeq
A density matrix of this form encompasses the case of a system in 
thermal equilibrium. We now imagine that all of the states are 
displaced by $\hat T(\bx,\bp)$ at time $t=0$. The resulting density 
matrix is then given by
\begin{align}
 \hat{\rho}(0)=\hat {T}(\bx,\bp)\hat \rho_0\hat {T}^{\dag}(\bx,\bp).
\label{Density_matrix}
\end{align}
The time evolution of this density matrix is given by 
\beq
 \hat{\rho}(t)=\hat \calU(t)\hat{\rho}(0)\hat \calU^\dag(t)
 =\hat \calU(t) \hat T(\bx,\bp) \hat{\rho}_0 \hat T^\dag(\bx,\bp) 
\hat \calU^\dag(t).
\eeq
Comparing Eqs.~(\ref{dystateII}) and (\ref{dynamicstate}), we 
have the operator identity
\beq
\hat \calU(t) \hat T(\bx,\bp) = e^{-\frac{i}{\hbar}S(t)} \hat
{T}(\bx(t),\bp(t)) e^{-i\hat H_0 t/\hbar }
\label{identity}
\eeq
where Eqs.~(\ref{xt}) and (\ref{pt}) define the time evolution of 
the displacement operator. We thus find
\beq
 \hat{\rho}(t) =\hat {T}(\bx(t),\bp(t))e^{-i\hat H_0 t/\hbar }\hat 
\rho_0e^{i\hat H_0 t/\hbar }\hat {T}^{\dag}(\bx(t),\bp(t)).
\eeq
This density matrix corresponds to each state $|\xi_k \ra$
evolving freely for a time $t$ and then being displaced along
the forced oscillator trajectory. If the states $|\xi_k\ra$ 
in $\hat \rho_0$ are in fact eigenstates $|\Psi_\alpha\ra$ of $\hat H_0$, we have 
the simpler result
\beq
\hat \rho(t)=\hat {T}(\bx(t),\bp(t))\hat \rho_0\hat 
{T}^{\dag}(\bx(t),\bp(t)).
\eeq
With this density matrix, the time-dependent density of 
the system is
\begin{align}
 n(\br,t)&={\rm Tr}[\hat \rho(t)\hat n(\br)] \nn \\
 &=\text{Tr}[\hat {T}(\bx(t),\bp(t))\hat \rho_0\hat 
{T}^{\dag}(\bx(t),\bp(t))\hat n(\br)] \nn \\
 &=\text{Tr}[\hat{\rho}_0\hat T^\dag(\bx(t),\bp(t))\hat{n}(\br)\hat 
T(\bx(t),\bp(t))].
 \end{align}
 Recalling that $\hat n(\br)=\sum_{i=1}^N\delta(\hat\br_i-\br)$ and 
using Eq.~(\ref{TfT}), we find 
 \begin{align}
 n(\br,t)&=\text{Tr}[\hat{\rho}_0\hat{n}(\br-\bx(t))] \nn \\
 &=n_0(\br-\bx (t)),
\end{align}
where $n_0(\br)={\rm Tr}[\hat \rho_0\hat n(\br)]$ is the density of the 
system before its displacement. We thus see that the density of the 
system experiences the same kind of rigid motion in the density matrix 
description as it does for a pure state.

In the rest of this section, we consider the special case in 
which the force is absent. We then find from Eq.~(\ref{dynamicstate})
that the dynamical state of the system at time $t$ is given by
\beq
|\Psi(t)\ra=\hat T\left (\bx_0(t),\bp_0(t)\right )
e^{-i\hat H_0 t/\hbar}|\Phi\ra.
\label{Psi_t_alpha}
\eeq
where $\bx_0(t)$ and $\bp_0(t)$ are given by Eqs.~(\ref{x0t}) and 
(\ref{p0t}). 
This result is of course consistent with the general result in
Eq.~(\ref{cm_osc}) for the dynamics of the centre of mass
coordinate. 
Eq.~(\ref{Psi_t_alpha}) also implies that the displacement
operator evolves according to 
\beq
\hat T\left ( \bx_0(t),\bp_0(t)\right )=e^{-i\hat H_0 t/\hbar }\hat 
T (\bx,\bp)e^{i\hat H_0 t/\hbar }
\label{emt}
\eeq
which is the force-free analogue of Eq.~(\ref{identity}).

 As we alluded to earlier, one can also understand the result 
in Eq.~(\ref{Psi_t_alpha}) from the perspective of coherent states. For 
simplicity, we take $\bx=\hat\bz z_0$ and $\bp=\hat \bz p_0$ in 
Eq.~(\ref{Psi_0_a}). Using Eq.~(\ref{a_cm}), the initial state in 
Eq.~(\ref{Psi_0_a}) can be written as
\beq
|\Psi(0)\ra=e^{-\gamma \hat a_{z}+\gamma^* \hat 
a^\dag_{z}}|\Psi_\alpha\ra,
\eeq
where 
$\gamma=z_0\sqrt{M\hbar\omega_z/2\hbar}-ip_0/\sqrt{2M\hbar\omega_z}$. 
We recognize this state as the analogue of a coherent state of a 
simple 
harmonic oscillator~\cite{Sakurai}. Its dynamics is then given by
\begin{align}
|\Psi(t)\ra&=e^{-i\hat H_0 t/\hbar}e^{-\gamma \hat a_{z}+\gamma^* \hat 
a^\dag_{z}}|\Psi_\alpha\ra \nn \\
&= e^{-iE_\alpha t/\hbar}e^{-\gamma \hat a_{z,\rm I}(-t)+\gamma^* \hat 
a^\dag_{z,\rm I}(-t)}|\Psi_\alpha\ra,
\label{psitco}
\end{align}
where 
\beq
\hat a_{z,\rm I}(t)\equiv e^{i\hat H_0 t/\hbar}\hat a_{z} e^{-i\hat 
H_0t/\hbar}=e^{-i\omega_z t}\hat a_{z}.
\label{azt}
\eeq
Substituting Eq.~(\ref{azt}) into Eq.~(\ref{psitco}) and using 
Eqs.~(\ref{a_cm}), we recover Eq.~(\ref{Psi_t_alpha}) where
$\bx_0(0)=\hat\bz z_0$ and $\bp_0(0)=\hat \bz p_0$.

Finally we discuss the experimental realization of the initial state in 
Eq.~(\ref{Psi_0_a}). The basic idea is to initiate the
oscillatory motion of a harmonically-confined system by a sudden
displacement of the trapping potential. To be specific, we take
the state of the undisplaced potential to be the eigenstate 
$|\Psi_\alpha\ra$. If the trapping potential is then displaced
in the $z$-direction by an amount $-z$, the state of the system
{\it relative to} the shifted potential is given by
\beq
|\Psi(0)\ra=\hat T\left ( 
\bx=\hat {\bf z}z,\bp=0\right )|\Psi_\alpha\ra.
\eeq
At a later time $t_0$, the state of the system according to
Eq.~(\ref{Psi_t_alpha}) is
\beq
|\Psi(t_0)\ra=e^{-\frac{i}{\hbar}E_\alpha t_0} \hat T\left ( 
\bx_0(t_0),\bp_0(t_0)\right )|\Psi_\alpha\ra,
\eeq
where, from Eqs.~(\ref{x0t}) and (\ref{p0t}),
$\bx_{0}(t_0) = z\cos \omega_z t_0\hat {\bf z}$ and $\bp_0(t_0)
= -M\omega_z z \sin \omega_z t_0\hat {\bf z}$. With an appropriate 
choice of
the initial displacement $z$ and time $t_0$, we can achieve the
initial conditions $\bx_{0}(t_0) = z_0\hat{\bf z}$ and
$\bp_0(t_0) =p_0 \hat{\bf z}$.
A more elaborate sequence of displacements of the trap in
different directions can in principle be used to achieve
arbitrary initial conditions.

\section{Dipole oscillations in the presence of
perturbations: extension of the HPT}
The theoretical development in this section is motivated by 
several recent experiments~\cite{Lye,Chen,Dries} which studied the 
centre of mass dynamics of trapped Bose condensates in the presence of 
a disorder potential. The disorder in these experiments is an 
example of an external perturbation which couples the centre of
mass and internal degrees of freedom. As a result of this
coupling, the energy associated with the centre of mass motion
is transferred to internal excitations, in other words,
mechanical energy is converted into `heat'. One can think of the
external perturbation as effectively exerting a drag
force on the centre of mass which leads to a damped oscillation.
Although the external potential acting on the system can be 
quite arbitrary, for
illustration purposes we will occasionally visualize it
as a disorder potential in the following.

A harmonically-confined system in the presence of an additional 
external potential $V_{\rm ext}(\br)$ is governed by the Hamiltonian
\beq
\hat H=\hat H_0 +\sum_{i=1}^NV_{\rm ext}(\hat \br_i) \equiv \hat
H_0 + \hat V_{\rm ext}.
\label{H_static}
\eeq
To investigate the dynamics of the centre of mass, we 
consider the Heisenberg equations of motion (the Heisenberg operators 
here are defined with respect to the full Hamiltonian $\hat H$)
\begin{align}
\label{eqRdis}
  \frac{d\hat{R}_{\mu}(t)}{dt}&=\frac{1}{i\hbar}[\hat{R}_\mu(t), \hat 
{H}]=\frac {\hat {P}_{\mu}(t)}{M}, \\
\frac{d\hat{P}_{\mu}(t)}{dt}&=\frac{1}{i\hbar}[\hat{P}_{\mu}(t),\hat{H}]
=-M\omega_\mu^2\hat{R}_\mu(t)+\hat F_\mu(t),
 \label{eqPdis}
\end{align}
where
\beq
\hat F_\mu=-\sum_{i=1}^N \frac{\partial V_{\rm ext}(\hat
\br_i)}{\partial \hat r_{i,\mu}}
\eeq
is the $\mu$-component of the external force operator. Equations 
(\ref{eqRdis}) and (\ref{eqPdis}) then lead to
\beq 
\frac{d^2\hat R_\mu }{ dt^2} + \omega_\mu^2 \hat R_\mu(t) = \frac{\hat 
F_\mu(t)}{ M}.
\label{RF}
\eeq
Taking the expectation value of both sides of Eq.~(\ref{RF}) with 
respect to the initial state $|\Psi(0)\ra$, we find that the 
$z$-component of the centre of mass position satisfies the equation
\beq 
\frac{d^2Z(t) }{ dt^2} + \omega_z^2 Z(t) = \frac{F(t)}{ M},
\label{cofmeq1}
\eeq
where $Z(t)=\langle \Psi(0)| \hat R_z(t) |\Psi(0) \rangle = \langle 
\Psi(t)| \hat R_z |\Psi(t) \rangle$ and
\beq 
F(t)=\langle \Psi(0)| \hat F_z(t) |\Psi(0) \rangle=\langle \Psi(t)| 
\hat F_z |\Psi(t) \rangle.
\label{F1}
\eeq
Eq.~(\ref{cofmeq1}) is analogous to Eq.~(\ref{cdoeq}) for the
forced oscillator considered earlier. However here, $F(t)$ is not
given explicitly but is defined by the dynamical state
$|\Psi(t)\ra$.

The dynamical evolution of the state $|\Psi(t)\ra$ is governed 
by the time-dependent Schr{\"o}dinger equation
\beq
i\hbar\frac{\pa}{\pa t}|\Psi(t)\ra=\hat H |\Psi(t)\ra.
\label{psi_evol}
\eeq
The physical situation of interest is one in which the dynamics
is initiated by suddenly shifting the harmonic trapping
potential at some instant of time as discussed at the end of the
previous section. For this reason, we will consider an
initial state of the form given in Eq.~(\ref{Psi_0_a}).
The state $|\Phi\ra$ being displaced,  will be specified later 
when we consider various experimental protocols for its
preparation. The following development does not depend on the
specific choice of the state $|\Phi\ra$.

In the interaction picture, the state $|\Psi_{\rm I}(t)\ra \equiv
e^{i\hat H_0 t/\hbar} |\Psi(t)\ra$ satisfies the equation
\beq
i\hbar\frac{\pa}{\pa t}|\Psi_{\rm I}(t)\ra=e^{i\hat H_0 t/\hbar} \hat
V_{\rm ext}e^{-i\hat H_0 t/\hbar} |\Psi_{\rm I}(t)\ra.
\label{psi_I_evol}
\eeq
We now define the state
\beq
|\tilde \Psi_{\rm I}(t)\ra = \hat{T}^\dag(\bx,\bp)|\Psi_{\rm I}(t)\ra
\eeq
which has the initial value $|\tilde \Psi_{\rm I}(0)\ra = |\Phi\ra$.
This state satisfies the equation
\begin{eqnarray}
i\hbar\frac{\pa}{\pa t}|\tilde \Psi_{\rm I}(t)\ra
&=& \hat{T}^\dag(\bx,\bp) e^{i\hat H_0 t/\hbar} \hat
V_{\rm ext}e^{-i\hat H_0 t/\hbar} \hat{T}(\bx,\bp) |\tilde
\Psi_{\rm I}(t)\ra
\nonumber \\
&=& e^{i\hat H_0 t/\hbar}  \hat{T}^\dag(\bx_0(t),\bp_0(t)) \hat
V_{\rm ext} \hat{T}(\bx_0(t),\bp_0(t)) e^{-i\hat H_0 t/\hbar}
|\tilde \Psi_{\rm I}(t)\ra \nonumber \\
&=& e^{i\hat H_0 t/\hbar} \sum_{i=1}^NV_{\rm ext}(\hat \br_i +
\bx_0(t)) e^{-i\hat H_0 t/\hbar}
|\tilde \Psi_{\rm I}(t)\ra.
\end{eqnarray}
To obtain this result we have used Eqs.~(\ref{emt}) and
(\ref{TfT}). The above equation can be interpreted as the
interaction picture evolution of the state $|\tilde \Psi(t)\ra =
e^{-i\hat H_0 t/\hbar} |\tilde \Psi_{\rm I}(t)\ra$ which satisfies the
equation
\beq
i\hbar\frac{\pa}{\pa t}|\tilde \Psi(t)\ra=\left (\hat H_0 +
\sum_{i=1}^NV_{\rm ext}(\hat \br_i +\bx_0(t))\right )|\tilde
\Psi(t)\ra\equiv \hat H(t)|\tilde \Psi(t)\ra ,
\label{tilde_psi_evol}
\eeq
with the initial condition $|\tilde \Psi(0)\ra = |\Phi\ra$. The
states $|\tilde \Psi(t)\ra$ and $|\Psi(t)\ra$ are related by
\beq
|\Psi(t)\ra = \hat{T}(\bx_0(t),\bp_0(t)) |\tilde \Psi(t)\ra.
\label{transf}
\eeq
We thus see that the state $|\Psi(t)\ra$ of interest, which evolves 
from $\hat{T}(\bx,\bp)|\Phi\ra$ according to the 
stationary Hamiltonian $\hat{H}$, 
can be obtained by a displacement of the state $|\tilde\Psi(t)\ra$ 
via Eq.~(\ref{transf}). The latter state corresponds to a different 
physical situation in which the system starts in the state
$|\Phi\ra$ and then evolves in the presence of
a {\it dynamic} potential oscillating at the trap frequency.
Equation (\ref{transf}) is the main result of this section. It
in fact reduces to the result given in Eq.~(\ref{Psi_t_alpha})
obtained in the context of the HPT when $\hat V_{\rm ext} \equiv
0$. For this reason, we refer to it as the {\it extended HPT}.
Its utility will become clear in the subsequent discussion.



\begin{figure}[!htbp]
\begin{center} 
 \includegraphics[angle=0, width=0.45\columnwidth]{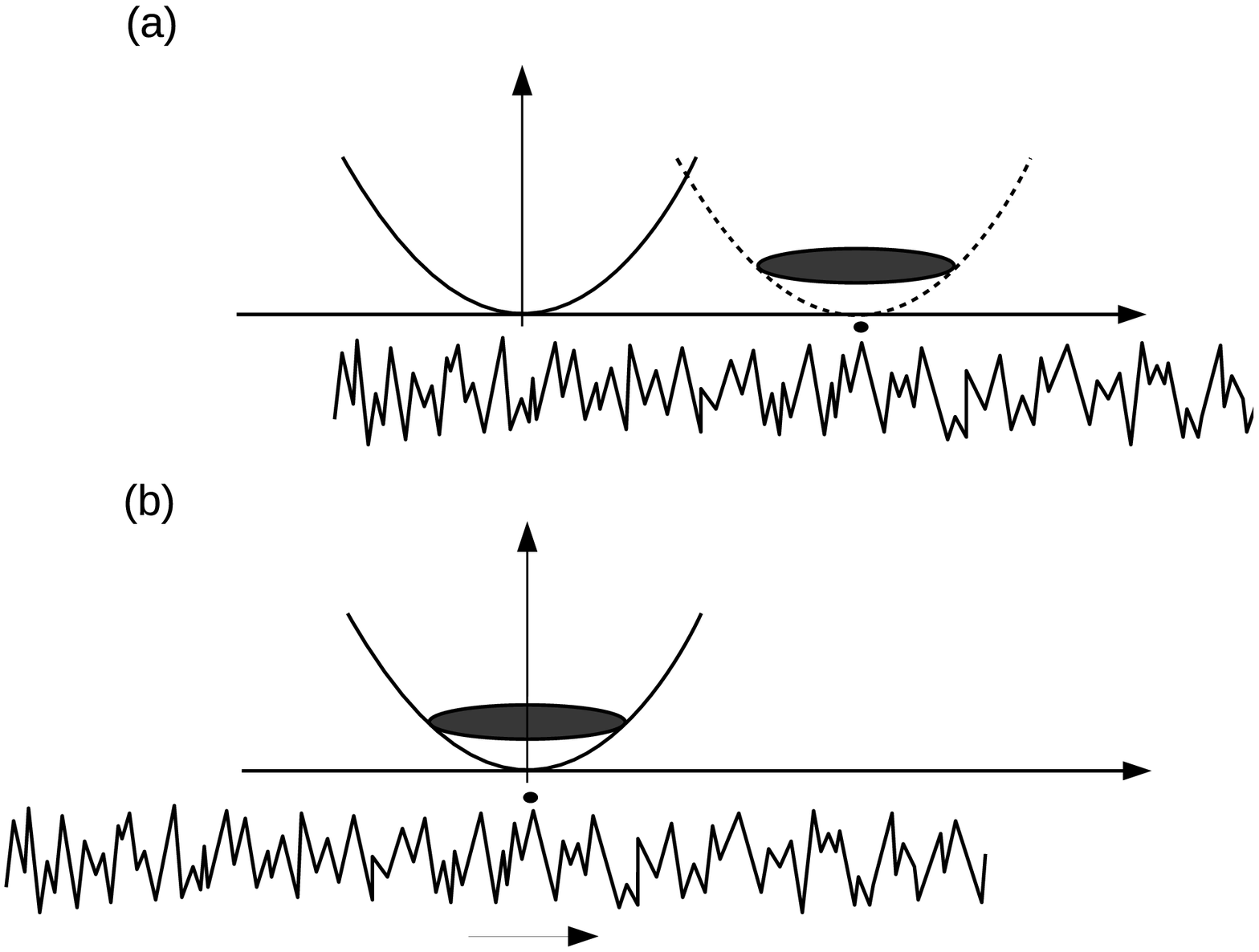}
 \includegraphics[angle=0, width=0.45\columnwidth]{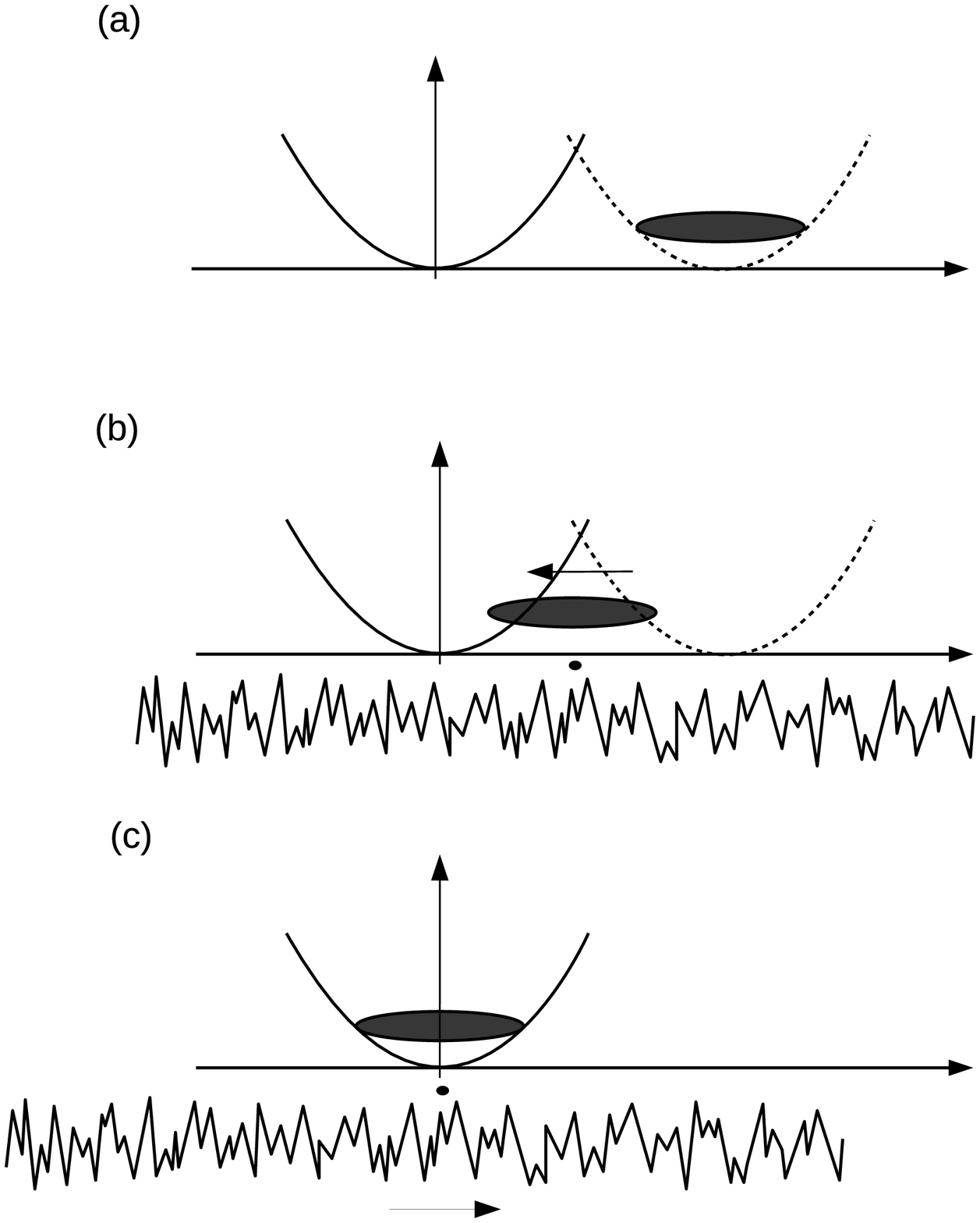}
 \caption
 {Left panel: (a) The condensate, originally in equilibrium with 
the unshifted trap (dashed) and the external potential, begins to 
oscillate about the centre of the shifted trap (solid). (b) The 
condensate, originally in equilibrium with the trap and external 
potential, is driven by an oscillating external potential. Right
panel: (a) At time $t<t_0$, the condensate is in its ground state with 
respect to the trapping potential indicated by the dashed curve. At 
$t=t_0<0$, the trapping potential is suddenly shifted to the origin, 
initiating the free oscillation of the condensate. (b) At $t=0$, the 
condensate arrives at $\bx=\hat \bz \cos\omega_z t_0$ with the velocity 
$\bv=\bp/M=\hat \bz \omega_z  z_0\sin\omega_z t_0$ and the external 
potential is suddenly switched on. For $t>0$, the condensate evolves 
according to the Hamiltonian in Eq.~(\ref{H_static}). (c) The external 
potential has a velocity $-\bv$ at $t=0$ and subsequently oscillates 
according to the free centre of mass motion. The condensate, initially 
stationary, begins to respond to the {\it dynamic} 
external potential.}
\label{oscillation}
\end{center}
\end{figure}

To give a concrete example of these general ideas, we consider
the effect of a disorder potential on the dynamics of a 
harmonically-confined Bose-condensed
gas. Fig.~\ref{oscillation} illustrates two possible protocols
for the initiation of the dynamics~\cite{Chen,Dries}.
In the left panel (a), we start with the condensate initially in its
ground state with respect to the trapping potential (dashed
curve) and the external (disorder) potential. At $t=0$, the
trapping potential is suddenly shifted to the origin (solid
curve) which initiates the centre of mass oscillation. The
initial state $|\Psi(0)\ra$ is the ground state of the condensate in
the total potential
\beq
V_{\rm trap}(\br - \bx) + V_{\rm ext}(\br),
\eeq
where $\bx = z_0 \hat \bz$. This state can be expressed as
\beq
|\Psi(0)\ra=\hat T(\bx,\bp=0)|\Phi_0\ra,
\label{Psi0}
\eeq
where $|\Phi_0\ra$ is the ground state in the
total potential
\beq
V_{\rm trap}(\br) + V_{\rm ext}(\br+\bx).
\eeq
This latter potential together with the state $|\Phi_0\ra$ 
is illustrated in (b) of the left panel.
The initial state $|\Phi_0\ra$ evolves into the state 
$|\tilde \Psi(t)\ra$ according to Eq.~(\ref{tilde_psi_evol}). 

In the situation illustrated in the right panel of
Fig.~\ref{oscillation}, the condensate starts off in the ground
state of the harmonic potential which is shifted to the origin
at some instant of time (a). The state is then allowed to evolve
freely for some interval of time after which the state of the
condensate is given by
\beq
|\Psi(0)\ra = \hat T(\bx,\bp) |\Phi_0\ra,
\eeq
where $|\Phi_0\ra$ is now the ground state of $\hat H_0$ (with 
the harmonic potential
$V_{\rm trap}(\br)$ centred on the origin) and the displacement 
operator $\hat T(\bx,\bp)$ determines the position and
velocity of the condensate at the time $t=0$ as shown in (b) of
the right panel of Fig.~\ref{oscillation}. At this instant, the
disorder potential is switched on and the system evolves
according to the Hamiltonian $\hat H$. 

We thus see that in both scenarios illustrated in
Fig.~\ref{oscillation}, the initial state can be expressed in
the form shown in Eq.~(\ref{Psi_0_a}). Although the two
initial states are different, the subsequent evolution
for $t>0$ (left panel (a) and right panel (b)) takes place in
both cases according to the Hamiltonian $\hat H$ to generate the
state $|\Psi(t)\ra$. This state is related to the state $|\tilde
\Psi(t)\ra$ through Eq.~(\ref{transf}).
This relationship implies that the force 
appearing in Eq.~(\ref{cofmeq1}) can be expressed as
\begin{align}
F(t) &= \langle \Psi(t)|\hat F_z | \Psi(t) \rangle \nn \\
&= \langle\tilde\Psi(t)|\hat T^\dag(\bx_0(t),\bp_0(t)) \hat F_{z}\hat 
T(\bx_0(t),\bp_0(t)) | \tilde \Psi(t)\rangle \nn \\
&= \langle\tilde\Psi(t)| \tilde F_{z}(t) | \tilde \Psi(t)\rangle \nn \\
&\equiv\tilde F(t),
\label{force_equ}
\end{align}
where we have used Eq.~(\ref{TfT}) to obtain the force operator
\beq
\tilde F_{z}(t) = -\sum_{i=1}^N \frac{\partial V_{\rm ext}(\hat \br_i + 
\bx(t))}{ \partial \hat
z_i}
\label{Ftilde}
\eeq
corresponding to the oscillating disorder potential. That the cloud 
experiences the same force due to the disorder in these two 
situations is by no means obvious and is a consequence of the 
validity of the extended HPT. This equivalence
was exploited in our earlier work~\cite{Wu} to determine the
disorder-induced damping in the limit of a weak disorder
potential. Since the state $|\tilde \Psi(t)\ra$ starts off at
$t=0$ in the
ground state $|\Phi_0\ra$ (either with or without
the disorder potential), the effect of a weak disorder potential
can be accounted for using conventional linear response theory.
On the other hand, the initial state in the $|\Psi(t)\ra$
dynamics is a highly excited state of the harmonic trapping
potential and conventional linear response theory cannot be
applied in this case.

The relationship in Eq.~(\ref{transf}) shows that there 
is an intimate connection between the two very distinct physical 
situations depicted in Fig.~\ref{oscillation} (left panel, (a)
or right panel, (b) and left panel, (b) or right panel, (c)). In
the first, the system starts in a highly excited state in which the
condensate is displaced from the minimum of the harmonic trap.
The energy of this state is given by
\beq
E(0) = \la\Psi(0)|\hat H |\Psi(0)\ra = \la\Phi_0|\hat 
T^\dagger(\bx,\bp)\hat H \hat T(\bx,\bp) |\Phi_0\ra.
\eeq
Using the properties of the displacement operator in
Eqs.~(\ref{TfT}) and (\ref{TfT2}), we have
\beq
\hat T^\dagger(\bx,\bp)\hat H \hat T(\bx,\bp) = \hat H_0
+\sum_{i=1}^N V_{\rm ext}(\br_i+\bx) + \frac{1}{M}\bp\cdot\hat
\bP + \sum_\mu M\omega_\mu^2 x_\mu \hat R_\mu + E_{\rm cm},
\eeq
where the centre of mass energy is defined as
\beq
E_{\rm cm} = \frac{p^2}{2M} + \frac{1}{2} \sum_\mu M\omega_\mu^2
x_\mu^2.
\eeq
Thus the initial energy is
\beq
E(0) = E_0 + E_{\rm cm} + \sum_\mu M\omega_\mu^2 x_\mu \la
\Phi_0|\hat R_\mu|\Phi_0\ra,
\eeq
with $E_0= \la \Phi_0|\hat H_0 +\sum_{i=1}^NV_{\rm
ext}(\br_i+\bx) |\Phi_0\ra$. We again note that $|\Phi_0\ra$
denotes different states for the two scenarios depicted
Fig.~\ref{oscillation}, but for both we have
$\la \Phi_0|\hat \bP|\Phi_0\ra = 0$.
The initial state then evolves in the presence of the static external 
potential according to Eq.~(\ref{psi_evol}) and during 
this evolution, the total energy of the system is conserved.

In the alternative point of view (Fig.~\ref{oscillation}: left panel, 
(b) or right panel, (c)), the condensate starts out near the
minimum of the harmonic trap and is 
driven by a {\it dynamic} external potential which oscillates at
the trap frequency. In view of Eq.~(\ref{transf}), the energy is 
given by
\begin{align}
\tilde E(t) &= \la \tilde \Psi(t) |\hat H(t) | \tilde \Psi(t)\ra
\nn\\
&=\la \Psi(t) | \hat T(\bx_0(t),\bp_0(t) \hat H(t) \hat
T^\dagger(\bx_0(t),\bp_0(t)| \Psi(t)\ra\nn\\
&= E(0) + E_{\rm cm} - \frac{1}{M} \bp_0(t) \cdot\la \Psi(t)|\hat \bP
| \Psi(t) \ra - M \sum_\mu \omega_\mu^2 x_{0\mu}(t) \la
\Psi(t)|\hat R_\mu | \Psi(t) \ra.
\label{tildeE}
\end{align}
Here, the dynamic 
perturbation continually  excites the condensate and the total energy 
increases as a function of time from the initial value $\tilde
E(0) = E_0$. Provided the external potential provides a coupling
between the centre of mass and internal degrees of freedom, one
would expect on physical grounds that the expectation value of the total
momentum $\hat \bP$ in the state $|\Psi(t)\ra$ should tend to zero
at long times. By the same token, the expectation value of $\hat
R_\mu$ should tend to some constant limiting value. In this
case, the time average of $\tilde E(t)$ tends to a finite
limiting value of $E(0)+E_{\rm cm}$. The important conclusion is
that the energy does not increase indefinitely as a result of
the dynamic perturbation and suggests that the system approaches
a steady state.
 
This interpretation of the long time behaviour is supported
by the time dependence of the centre of mass coordinate.
In view of 
Eq.~(\ref{transf}), we have 
\begin{align}
\tilde Z(t)&\equiv\la\tilde\Psi(t)|\hat R_z|\tilde\Psi(t)\ra \nn \\
&=\la\Psi(t)|\hat T(\bx_0(t),\bp_0(t))\hat R_z\hat 
T^\dag(\bx_0(t),\bp_0(t))|\Psi(t)\ra \nn \\
&=Z(t)-x_{0z}(t),
\end{align}
where $x_{0z}(t)$ is given by Eq.~(\ref{x0t}).
In Fig.~\ref{oscillation} (left panel (b) or right panel (c)) we
have placed a marker (filled dot) tied to the external
potential that initially coincides with the origin. The position 
of this marker is $z_M(t) = x_{0z}(0) - x_{0z}(t)$ and the
position of the centre of mass relative to it is $\tilde
Z(t)-z_M(t) = Z(t) - x_{0z}(0)$. With the assumption that
$\lim_{t\rightarrow \infty}Z(t)=Z_\infty$, we thus conclude that the 
condensate in Fig.~\ref{oscillation} (left panel (b) or right
panel (c)) moves synchronously 
with the the external potential at long times with the
centre of mass located at $Z_\infty - x_{0z}(0)$ relative to
the marker.

The final position $Z_\infty$ of the centre of mass in
Fig.~\ref{oscillation} (left panel (a) or right panel (b))
depends on the details of the external perturbation. For the
example of a {\it weak} disorder potential, one expects
$|\Psi(t)\ra$ to approach a 
quasi-equilibrium state~\footnote{At long times 
the quantum state $|\Psi(t)\ra$ is a highly excited state which is a 
complex superposition of all eigenstates of $\hat H$. However, 
physically we expect that this state is essentially equivalent to a 
thermal equilibrium state in the sense that the expectation values of 
various macroscopic  physical quantities obtained using the dynamical 
state will be very similar to the averages calculated using a thermal 
equilibrium ensemble.}
which is centred on the minimum of the
harmonic potential, i.e., $Z_\infty \simeq 0$. However, if the
disorder is very strong, the initial state can be localized by
the disorder potential~\cite{Chen} and $Z_\infty \simeq x_{0z}(0)$. In this
case, the centre of mass is pinned to the position of the marker
in Fig.~\ref{oscillation} (left panel (b) or right panel (c)) so
that $\tilde Z(t) \simeq z_M(t)$. This motion gives a centre of
mass energy $M\omega_z^2 z_M(t)^2/2+M\dot z_M(t)^2/2 = 2E_{{\rm
cm},z} - M\omega_z^2x_{0z}(0)x_{0z}(t)$. The oscillatory term in
this result corresponds to the $\mu =z$ contribution from the last
term in Eq.~(\ref{tildeE}). Its time average is of course zero.

We conclude this section with a few general comments. The
dissipation of the harmonically-confined condensate in the
presence of an external perturbation is analogous to the
dissipation experienced by a uniform superfluid moving past 
an impurity with a {\it constant} velocity. Galilean
invariance allows one to consider the latter situation from the
equivalent point of view of the impurity moving with a {\it
constant} velocity through a stationary
superfluid~\cite{Lifshitz}. Although a
Galilean transformation does not apply to the harmonically
trapped gas, the situation depicted in Fig.~\ref{oscillation}
(left panel (b) or right panel (c)) is analogous to the moving
impurity in that the external (disorder) potential is moving relative 
to the gas which, at least initially, is stationary. However, it
would be incorrect to think of these situations as having arisen
by means of a transformation
to a frame of reference in which the centre of mass, say, is at
rest. Such a frame would not oscillate freely at the trap
frequency as the external potential is required to do according 
to the extended HPT.

Although the extended HPT was motivated by other considerations,
it is worth pointing out that Eqs.~(\ref{psi_evol}) and
(\ref{tilde_psi_evol}) are in fact related by means of a
coordinate transformation. We interpret the state vector
$|\tilde \Psi(t)\ra$ satisfying Eq.~(\ref{tilde_psi_evol}) as the
state in the `laboratory' frame of reference. We now imagine
making a transformation to the non-inertial frame of reference
in which $V_{\rm ext}(\br+\bx_0(t))$ is stationary~\cite{Takagi}. 
In this frame
of reference, the trapping potential is of course non-stationary.
Nevertheless, when the harmonic trapping potential is
transformed, one finds that the state vector in the non-inertial
frame is the solution of Eq.~(\ref{psi_evol}) {\it provided}
$\bx_0(t)$ is given by Eq.~(\ref{x0t}). It is only in this
circumstance that the non-inertial forces generated by the
transformation from the laboratory to the non-inertial frame 
are eliminated, with the evolution of the state vector in the
non-inertial frame being governed simply by $\hat H$. That
Eq.~(\ref{transf}) provides the
relation between the state vectors in the two frames of
reference is a direct
consequence of the fact that the trapping potential is harmonic.

\section{Harmonically confined systems in the presence of an 
oscillating external potential}
In the course of the derivation of the extended HPT, we
encountered the state $|\tilde \Psi(t)\ra$ which satisfies the
Schr\"odinger equation (\ref{tilde_psi_evol}). This state
begins in the state $|\Phi\ra$ at $t=0$ and is subjected to a
dynamic perturbing potential. One
would usually expect a dynamic perturbation of this kind to lead to
a continuous energy absorption, but as discussed earlier, 
for the special case in which the
perturbation oscillates at the trap frequency, the time-averaged
energy absorption rate eventually goes to zero. This would {\it not}
be the case if the perturbing potential oscillated at some
arbitrary frequency. In this section, we examine this more
general situation. As a concrete example, one might consider 
the effect of an oscillating optical lattice on a 
harmonically-confined system.
Oscillations of the lattice in a certain range of frequency can lead to 
suppression of the tunnelling and the so-called dynamically induced 
phase transition~\cite{Eckardt,Lignier}. Here, we show that a
calculation of the energy absorption or centre of mass position
provides a 
probe of the optical conductivity of the system when the amplitude
of the oscillating perturbing potential is small. This analysis
extends the results of earlier work~\cite{Tokuno} to the case of a
{\it harmonically-confined} system subjected to
an {\it arbitrary} oscillating external potential. 

The physical problem of interest is described by the Hamiltonian
\beq
\hat H(t)=\hat H_0 +\sum_{i=1}^NV_{\rm ext}(\hat \br_i -
\br_0(t)),
\label{H_shaking}
\eeq
where $\br_0(t)$ is an arbitrary, time-dependent displacement
vector. We refer to this kind of external potential as a
`shaking' perturbation. If $\br_0(t)$ is small on the length
scale of variations of the external potential,
the Hamiltonian can be expanded as
\beq
\hat H(t)\simeq \hat H + \hat H'(t),
\label{H_H'}
\eeq
where $\hat H$ is given in Eq.~(\ref{H_static}) and the perturbation is
\beq
\hat H'(t)= -\sum_{i=1}^N \nabla V_{\rm ext}(\hat \br_i)\cdot
\br_0(t) = -\int d\br \nabla V_{\rm ext}(\br)\cdot\br_0(t) \hat n(\br).
\label{H'}
\eeq
Although the perturbation is seen to couple to the density operator, 
the coupling has the very special form of the gradient of the
external potential. This allows the perturbation to be expressed
in terms of the commutator relation
\beq
\hat H'(t)= \sum_\mu r_{0\mu}(t) \frac{m}{i\hbar}
[\hat J_\mu,\sum_{i=1}^N V_{\rm ext}(\hat \br_i)],
\eeq
where $\hat J_\mu = \hat P_\mu/m$ is the total current operator.
Expressing $\sum_{i=1}^N V_{\rm ext}(\hat \br_i)$ in terms of
$\hat H$ and noting that $\hat J_\mu$ commutes with the total
kinetic energy and interatomic interactions, we find
\beq
\hat H'(t)= 
\sum_\mu r_{0\mu}(t) \frac{m}{i\hbar}[\hat J_\mu,\hat
H]+Nm\sum_\mu \omega_\mu^2 r_{0\mu}(t)\hat R_\mu.
\eeq
The second term on the right hand side comes from the commutator
of $\hat J_\mu$ and $\sum_{i=1}^N V_{\rm tr}(\hat \br_i)$. This
form of the perturbation provides a direct route to the current
response of the system. In the following, it is convenient to 
write the perturbation as
\beq
\hat H'(t) \equiv \sum_{i\mu}r_{0\mu}(t) \hat A_{i\mu},
\eeq
where
\beq
\label{opA1}
\hat A_{1\mu}= \frac{m}{i\hbar}[\hat J_\mu,\hat H],
\eeq
and
\beq
\hat A_{2\mu}=  Nm\omega_\mu^2\hat R_\mu .
\label{opA2}
\eeq

The total energy of the system in the presence of the
perturbation is $\tilde E(t) = \la \tilde \Psi(t)|
\hat H(t) | \tilde \Psi(t)\ra$, where $| \tilde \Psi(t)\ra$ 
is the state of the system at time $t$. The tilde notation is
used since the Hamiltonian in Eq.~(\ref{H_shaking}) is analogous
to the Hamiltonian in Eq.~(\ref{tilde_psi_evol}).
The rate of energy absorption is given quite generally by
\beq
\frac{d\tilde  E}{dt} = \la \tilde \Psi(t)| \frac{d\hat
H'(t)}{dt} | \tilde \Psi(t)\ra
\eeq
If we assume that the perturbation is turned on at $t= t_0$ and
that the system starts out in the ground state $|\Phi_0\ra$ of
$\hat H$, the energy absorption rate in the linear response
regime is given by
\begin{align}
 \frac{d\tilde E}{dt}&= \sum_{i\mu} \dot r_{i\mu}(t)\la \Phi_0|\hat A_{i\mu}|\Phi_0\ra -
\sum_{i\mu,j\nu}\dot r_{0\mu}(t) \int_{-\infty}^\infty d t' 
\chi_{i\mu.j\nu}(t-t') r_{0\nu}(t')
\label{dE/dt}
\end{align}
where we have defined the retarded response functions
\begin{align}
 \chi_{i\mu,j\nu}(t-t')\equiv 
\frac{i}{\hbar}\theta(t-t')\la\Phi_0|[\hat A_{i\mu}(t),\hat 
A_{j\nu}(t')]|\Phi_0\ra.
\end{align}
The Heisenberg operators
appearing in the response function are
$\hat A_{i\mu}(t) = e^{i\hat Ht/\hbar} \hat A_{i\mu}
e^{-i\hat Ht/\hbar}$. By introducing the correlation function
\beq
\phi_{i\mu,j\nu}(t-t') \equiv \frac{1}{2\hbar}\la\Phi_0|[\hat
A_{i\mu}(t),\hat A_{j\nu}(t')]|\Phi_0\ra,
\label{phi_t}
\eeq
the response function can be written as
\beq
 \chi_{i\mu,j\nu}(t-t')= 2i\theta(t-t')
\phi_{i\mu,j\nu}(t-t').
\eeq

We now consider the energy absorption rate for the special 
case of a monochromatic displacement, namely $r_{0\mu}(t) = 
\theta(t-t_0)(r_{0\mu}e^{-i\omega t} + r_{0\mu}^*e^{i\omega t})/2$. 
Taking the limit $t_0 \to -\infty$ and averaging Eq.~(\ref{dE/dt})
over one period $T=2\pi/\omega$, we find
\begin{align}
\overline{\frac{d\tilde E}{dt}}&\equiv \frac{1}{T}\int_0^T
\frac{d\tilde E}{dt} \nn \\
&= \frac{\omega}{2} \sum_{i\mu,j\nu} r_{0\mu}^*
\phi_{i\mu,j\nu}(\omega)r_{0\nu},
\label{edrinter}
\end{align}
where $\phi_{i\mu,j\nu}(\omega)$, the Fourier transform of
Eq.~(\ref{phi_t}), has the spectral representation (here, $\hat
H|\Phi_m\ra = E_m|\Phi_m\ra$)
\beq
\phi_{i\mu,j\nu}(\omega)=\pi\sum_n\left \{\la
\Phi_0|\hat A_{i\mu}|\Phi_n\ra\la
\Phi_n|\hat A_{j\nu}|\Phi_0\ra\delta(\hbar\omega-E_{n 
0})- \la \Phi_0|\hat A_{j\nu}|\Phi_n\ra\la \Phi_n|\hat
A_{i\mu}|\Phi_0\ra\delta(\hbar\omega+E_{n 0})\right \}.
\label{phiab}
\eeq
Since $\phi_{i\mu,j\nu}^*(\omega) = \phi_{j\nu,i\mu}(\omega)$,
it is clear that Eq.~(\ref{edrinter}) is manifestly real.
Furthermore, time-reversal symmetry implies that
$\phi_{i\mu,j\nu}(\omega) = \phi_{j\nu,i\mu}(\omega)$ and hence
that $\phi_{i\mu,j\nu}(\omega)$ is real. In this case, ${\rm
Im}\chi_{i\mu,j\nu}(\omega) = \phi_{i\mu,j\nu}(\omega)$ and
${\rm Re} \chi_{i\mu,j\nu}(\omega)$ is obtained from
$\phi_{i\mu,j\nu}(\omega)$ by a Kramers-Kronig
relation~\cite{Lifshitz}.

From Eqs.~(\ref{opA1}) and (\ref{opA2}), we find
\beq
\la\Phi_m|\hat A_{1\mu}|\Phi_n\ra=\frac{m}{i\hbar}E_{nm}\la\Phi_m|\hat 
J_\mu|\Phi_n\ra
\label{Amele}
\eeq
and
\beq
\la\Phi_m|\hat A_{2\mu}|\Phi_n\ra
=\frac{i\hbar m\omega_\mu^2}{E_{nm}}\la\Phi_m|\hat
J_\mu|\Phi_n\ra.
\label{Bmele}
\eeq
The latter equation follows from the identity
\beq
[\hat R_{\mu},\hat H]=\frac{i\hbar}{N}\hat J_{\mu}.
\eeq
In the above equations, $E_{nm} = E_n -E_m$. We see that all the
required matrix elements can be expressed in terms of those of
the total current operator.

Using Eqs.~(\ref{Amele}) and~(\ref{Bmele}) in Eq.~(\ref{phiab}), we 
obtain
\begin{align}
\phi_{1\mu,1\nu}(\omega)&=m^2\omega^2{\rm Im}\Pi_{\mu\nu}(\omega),
\label{phi11}
\\
\phi_{1\mu,2\nu}(\omega)&= -m^2\omega_\nu^2{\rm Im}\Pi_{\mu\nu}(\omega),
\label{phi12}
\\
\phi_{2\mu,1\nu}(\omega)&= -m^2\omega_\mu^2{\rm Im}\Pi_{\mu\nu}(\omega),
\label{phi21}
\\
\phi_{2\mu,2\nu}(\omega)&=m^2\frac{\omega_\mu^2\omega_\nu^2}{\omega^2}{\rm 
Im}\Pi_{\mu\nu}(\omega),
\label{phi22}
\end{align}
where
\beq
{\rm Im}\Pi_{\mu\nu}(\omega)=\pi\sum_n\left \{\la
\Phi_0|\hat J_{\mu}|\Phi_n\ra\la
\Phi_n|\hat J_{\nu}|\Phi_0\ra\delta(\hbar\omega-E_{n 
0})- \la \Phi_0|\hat J_{\nu}|\Phi_n\ra\la \Phi_n|\hat
J_{\mu}|\Phi_0\ra\delta(\hbar\omega+E_{n 0})\right \}.
\label{Imchiab}
\eeq
This quantity is the imaginary part of the Fourier transform of
the current-current response function
\beq
\Pi_{\mu\nu}(t-t')=\frac{i}{\hbar}\theta(t-t')\la \Phi_0|[\hat 
J_{\mu}(t),\hat J_{\nu}(t')]|\Phi_0\ra.
\label{Current_corr_1}
\eeq
It defines the real part of the optical conductivity according to~\cite{Mahan}
\beq
{\rm Re}\sigma_{\mu\nu}(\omega)=\frac{1}{\omega}{\rm Im}
\Pi_{\mu\nu}(\omega).
\eeq
We emphasize that this is the optical conductivity of
the system described by the Hamiltonian $\hat H$ in
Eq.~(\ref{H_static}) which includes both the harmonic and external
potentials.

Inserting Eqs.~(\ref{phi11})-(\ref{phi22}) into
Eq.~(\ref{edrinter}), we thus find
\begin{align}
\overline{\frac{d\tilde E}{dt}}=\frac{m^2\omega^3}{2}\sum_{\mu\nu}
r_{0\mu}^*r_{0\nu}\bigg
(1-\frac{\omega_\mu^2}{\omega^2}\bigg )
\left ( 1-\frac{\omega_\nu^2}{\omega^2}\right ){\rm Im} \Pi_{\mu\nu}(\omega).
\label{edrf}
\end{align}
This is a general result valid for any
harmonically-confined system in the presence of an arbitrary
oscillating external potential. 
We thus see that the time-averaged energy absorption rate in the
presence of harmonic confinement is proportional to the optical
conductivity.
If the oscillation is restricted
to the $z$-direction, i.e. $\br_0(t) = z_0 \hat {\bf z} \sin
\omega t$, the energy absorption rate becomes
\beq
\overline{\frac{d\tilde E}{dt}}=\frac{1}{2}m^2\omega^3z_0^2
\left (1-\frac{\omega_z^2}{\omega^2}\right )^2
{\rm Im} \Pi_{zz}(\omega).
\label{ear}
\eeq
In the absence of harmonic confinement ($\omega_z\rightarrow
0$), Eq.~(\ref{ear}) reduces to the result given in Ref.~\cite{Tokuno}
which was derived for the special case of an oscillating 
uniform optical lattice. An alternative derivation of this
limiting result is provided in Appendix A; this derivation
points out a shortcoming of the original derivation in
Ref.~\cite{Tokuno}.

It is interesting to observe that the energy absorption rate in
Eq.~(\ref{ear}) vanishes if the external perturbation is 
oscillating at the 
trapping frequency $\omega=\omega_z$. This in fact is an exact
result and is not a consequence of the perturbative analysis. If
we take $\br_0(t) = x_{0z}(t) \hat \bz$ where $x_{0z}(t)$ is
given in Eq.~(\ref{x0t}), the Hamiltonian in
Eq.~(\ref{H_shaking}) takes the form of the Hamiltonian in
Eq.~(\ref{tilde_psi_evol}). According to the extended HPT, the
time-averaged energy absorption rate goes to zero at long times
(see Eq.~(\ref{tildeE}) and the discussion thereafter). This
result applies for any amplitude of the oscillating potential
and in particular, accounts for the second order result in
Eq.~(\ref{ear}). An alternative proof is given in Appendix B
where we consider the oscillation of the external 
potential to be turned on {\it adiabatically}. If the
oscillation occurs at the frequency $\omega=\omega_z$, we show
that the
initial state of the system, $|\Phi_0\ra$, evolves into a state
which moves together with the oscillating potential. This
dynamical state has a vanishing time-averaged energy absorption
rate which is consistent with the $\omega=\omega_z$ result in
Eq.~(\ref{ear}).

Eq.~(\ref{edrf}) or (\ref{ear}) shows that information about 
the current-current correlation function can in principle be 
accessed via the energy absorption rate. However, the latter is
not a quantity that is easily measured. On the other hand, a
measurement of the centre of mass motion is relatively
straightforward and provides an alternative
means of probing the current-current correlation function.
For simplicity, we consider a perturbation shaking in the
$z$-direction. The relevant centre of mass coordinate is 
$Z(t) \equiv \la\Psi(t)|\hat R_z|\Psi(t)\ra$ which, in linear
response, is given by
\begin{align}
Z(t)&=\frac{1}{Nm\omega_z^2}\sum_{j}\int_{-\infty}^\infty
d t' \chi_{2z,jz}(t-t') z_0\sin\omega t'\nn \\
&=\frac{z_0}{Nm\omega_z^2}\bigg [ 
\sin\omega t  \sum_{j} {\rm Re}\chi_{2z,jz}(\omega)
-\cos\omega t \sum_{j} {\rm Im}\chi_{2z,jz}(\omega) \bigg ].
\label{comtr}
\end{align}
We see that $Z(t)$ oscillates at the frequency $\omega$ with
some phase lag relative to the oscillating external potential. If
we compare this with the expected experimental centre of mass
trajectory $Z_{\rm exp}(t)=Z_0 \sin(\omega t -\varphi)$, we see
immediately that
\beq
\sum_j {\rm Im}\chi_{2z,jz}(\omega)
= \frac{Z_0 }{z_0}Nm\omega_z^2\sin\varphi.
\label{Imsum}
\eeq
Using Eqs.~(\ref{phi21}) and (\ref{phi22}), we find
\beq
{\rm Im}\Pi_{zz}(\omega) = \frac{Z_0 }{z_0}
\frac{N\sin\varphi}{m(\omega_z^2/\omega^2-1)}.
\eeq
This result indicates that the current-current correlation function
can be obtained experimentally through a measurement of the
centre of mass oscillation amplitude $Z_0$ and phase lag
$\varphi$. This kind of measurement is thus a direct probe
of the optical conductivity.

\section{Conclusions}
In this paper we have studied 
various aspects of the dynamics of harmonically-confined 
atomic systems. The results we have obtained are of a general
nature and have a broad applicability to trapped atomic gases.
We have focused, in particular, on the effect that various
external perturbations have on the dynamical evolution of the 
many-body wavefunction. In the case of
a perturbation that couples solely to the
centre of mass, we were able to obtain an
explicit expression for the Schr{\"o}dinger
evolution operator; Dobson's Harmonic
Potential Theorem follows naturally from this
result. 

For more general external perturbations, the centre of
mass and internal degrees of freedom are coupled and dissipation
of the centre of mass motion sets in. We have proved an
extension of the HPT which demonstrates that 
this dissipative dynamics can be considered from two distinct
points of view. In the first, the evolution of an initial
nonequilibrim state takes place in the
presence of a {\it static} external potential. On the other hand,
one can equivalently think of the evolution as taking place in
the presence of an external potential that itself moves according to
the trajectory of a harmonically-confined particle. Here,
the trapped atomic cloud starts off in an
initial equilibrium state and is then continually
excited by a {\it dynamic} perturbation. This latter point of
view has the advantage that the calculation of the damping of
the centre of mass motion can be addressed by means of linear 
response theory when the perturbation is weak~\cite{Wu}. 

We next considered the response of a harmonically-confined
system to a `shaking' potential. For a weak perturbation, this
response is directly related to
current-current correlations and hence the optical conductivity.
Our result for the energy absorption is a generalization of one
obtained previously~\cite{Tokuno}. We have also shown
that the optical conductivity can be probed by measuring the
trajectory of the centre of mass itself. This may in fact be the
most feasible way of determining the optical conductivity
experimentally.

\section*{acknowledgments}
This work was supported by a grant from the Natural Sciences and
Engineering Research Council of Canada.

\appendix
\section{}
In this appendix we give a different derivation of the result
 in Eq.~(\ref{ear}) for the special case of no harmonic
confinement ($\omega_\mu = 0$).
The Hamiltonian in this case is
\beq
\hat H(t)=\sum_{i=1}^N\left [\frac{\hat \bp_i^2}{2m} 
+V_{\rm ext}(\hat x_i,\hat y_i,\hat z_i-z_0(t))\right
]+\sum_{i<j} v(\hat \br_i-\hat \br_j),
\label{H(t)}
\eeq
where we allow the displacement in the $z$-direction to have an
arbitrary time dependence. A Hamiltonian of this kind was
considered in Ref.~\cite{Tokuno} for the case in which $V_{\rm ext}$
corresponds to a uniform optical lattice. In this context, the
displacement $z_0(t)$ provides a phase modulation of the lattice
potential.

We observe that the Hamiltonian can be expressed as
\beq
\hat H(t)=\hat U^\dag(t) \hat H \hat U (t),
\label{phUHU}
\eeq
where $\hat H$ is the Hamiltonian in Eq.~(\ref{H(t)}) with
$z_0(t)\equiv 0$, and $\hat U(t)$ is the translation operator
\beq
\hat U (t) =\exp\left \{i z_0(t)\hat P_z/\hbar \right \}
=\exp\left \{imz_0(t)\hat J_z/\hbar\right \}.
\label{transoperator}
\eeq
The dynamic state of the system evolves according 
to the Schr{\"o}dinger equation
\begin{align}
i\hbar\frac{d |\Psi(t)\ra}{dt}&=\hat H(t) |\Psi(t)\ra \nn \\
&=\hat U^\dag (t)\hat H \hat U(t)|\Psi(t)\ra.
\end{align}
Defining the state
\beq
|\tilde \Psi (t)\ra=\hat U (t) |\Psi(t)\ra,
\eeq
we find that $|\tilde \Psi(t)\ra$ satisfies the equation
\begin{align}
i\hbar \frac{d|\tilde \Psi(t)\ra}{dt} =\tilde H (t) |\tilde \Psi(t)\ra,
\label{schtildepsi}
\end{align}
where 
\beq
\tilde H (t) =\hat H -m\dot z_0(t) \hat J_z.
\label{ap5tildeH}
\eeq
We see that the Hamiltonian governing the evolution 
of the state $|\tilde \Psi(t)\ra$ contains a perturbation 
proportional to the total current operator. It should be
emphasized that $|\tilde \Psi(t)\ra$ is {\it not} the state
of the system as seen in the non-inertial frame of 
reference in which the external potential is stationary. To
obtain the state in this frame of reference one must apply a
momentum boost in addition to the spatial displacement provided
by $\hat U(t)$~\cite{Takagi}.

The total energy of the system is given by
\begin{align}
E(t)&=\la \Psi(t)|\hat H(t)|\Psi(t)\ra \nn \\
&=\la \tilde \Psi(t) |\hat H |\tilde \Psi(t)\ra.
\label{aveenergy}
\end{align}
Using Eq.~(\ref{schtildepsi}) and Eq.~(\ref{aveenergy}),
 we find that the energy absorption rate is given by
\begin{align}
\frac{dE}{dt}&=\frac{1}{i\hbar}\la \tilde \Psi(t) |
[\hat H, \tilde H (t)]|\tilde \Psi(t)\ra \nn \\
&= -\frac{m\dot z_0(t)}{i\hbar}\la \tilde \Psi(t) |
[\hat H, \hat J_z]|\tilde \Psi(t)\ra.
\end{align}
Introducing the interaction picture state vector
$|\tilde \Psi_{\rm I}(t)\ra\equiv \exp(i\hat H t/\hbar)
|\tilde \Psi(t)\ra$, we have
\begin{align}
\frac{dE}{dt}&=  -\frac{m\dot z_0(t)}{i\hbar}\la \tilde \Psi_{\rm I}(t) |
[\hat H, \hat J_z(t)]|\tilde \Psi_{\rm I}(t)\ra,
\label{ER}
\end{align}
where $\hat J_z(t)\equiv 
\exp(i\hat H t/\hbar)\hat J_z \exp(-i\hat H t/\hbar)$.
The state $|\tilde \Psi_{\rm I}(t)\ra$ evolves according to 
\beq
i\hbar \frac{d|\tilde \Psi_{\rm I}(t)\ra}{dt}=-m \dot z_0(t)
\hat J_z(t)|\tilde \Psi_{\rm I}(t)\ra.
\eeq
First order perturbation theory gives 
\beq
|\tilde \Psi_{\rm I}(t)\ra\simeq |\Phi_0\ra -\frac{m}{i\hbar}
\int_{t_0}^t dt' \dot z_0(t') \hat J_z(t') |\Phi_0\ra,
\label{tPsi}
\eeq
where we assume that $z_0(t) \equiv 0$ for $t\le t_0$ and that
$|\Phi_0\ra$ is the ground state of $\hat H$.
Substituting Eq. (\ref{tPsi}) into Eq. (\ref {ER}) we have 
\begin{align}
\frac{dE}{dt}=&-\frac{m}{i\hbar}\dot z_0(t)\la \Phi_0 |
[\hat H, \hat J_z(t)]| \Phi_0\ra \nn \\
&+ \left (\frac{m}{i\hbar}\right )^2 \dot z_0(t)\int_{t_0}^t 
dt' \dot z_0(t')\la \Phi_0|[[\hat H,\hat J_z(t)],
\hat J_z(t')]|\Phi_0\ra.
\label{edrphasemn}
\end{align}
The first term on the right hand side of this equation 
vanishes since $|\Phi_0\ra$ is the ground state of 
$\hat H$. Using 
\beq
i\hbar \frac{d}{dt}\hat J_z(t)=[\hat J_z(t),
\hat H],
\eeq
Eq.~(\ref{edrphasemn}) can be written as
\begin{align}
\frac{dE}{dt}&=-\frac{m^2}{i\hbar}\dot z_0(t)\int_{t_0}^t dt' 
\dot z_0(t')\frac{\pa}{\pa t}\la \Phi_0|[\hat J_z(t),
\hat J_z(t')]|\Phi_0\ra\nn\\
& =m^2\dot z_0(t)\frac{d}{dt}
\int_{-\infty}^\infty dt' \Pi_{zz}(t-t')\dot z_0(t'),
\label{edrpmsys}
\end{align}
where $\Pi_{zz}(t-t')$ is defined in Eq.~(\ref{Current_corr_1}).
This is a general result for any displacement $z_0(t)$ that
vanishes for $t\le t_0$.

For the case of a sinusoidal perturbation, $z_0(t) = z_0 \sin
\omega t$, we can take the limit $t_0 \to -\infty$ and obtain
\beq
\frac{dE}{dt} =m^2z_0^2\omega^3\cos\omega t\left 
[-\sin\omega t\, {\rm Re}\Pi_{zz} (\omega)
+\cos\omega t \,{\rm Im}\Pi_{zz}(\omega)\right ].
\label{edrpmsin}
\eeq
Averaging this expression over one period, we obtain
\begin{align}
\overline{\frac{d E}{dt}}&=\frac{1}{T}\int_0^T\frac{dE}{dt} dt \nn \\
&=\frac{1}{2}m^2z_0^2\omega^3{\rm Im}\Pi_{zz}(\omega),
\label{EA}
\end{align}
which is the result given in Eq.~(\ref{ear}) for the case of
$\omega_z=0$.

We now point out that the above is {\it not} in fact the
derivation given in Ref.~\cite{Tokuno}. Instead of the correct
expression for the energy given in Eq.~(\ref{aveenergy}), the 
authors of Ref.~\cite{Tokuno} take the energy of the system to
be $\tilde E(t)=\la\tilde\Psi(t)|\tilde H (t)|\tilde
\Psi(t)\ra$, where $\tilde\Psi(t)$ is the solution of
Eq.~(\ref{schtildepsi}). The
energy absorption rate in Ref.~\cite{Tokuno} is then defined to
be
\begin{align}
\frac{d\tilde E}{dt}=\la \tilde \Psi(t)|\frac{\pa\tilde H (t)}{\pa t}
|\tilde \Psi(t)\ra.
\label{edrtokuno}
\end{align}
With $\tilde H(t)$ given by Eq.~(\ref{ap5tildeH}), one has
\begin{align}
\frac{d\tilde E}{dt}&=-m\ddot z_0(t)\la \tilde \Psi(t)|\hat J_{z} 
|\tilde \Psi(t)\ra \nn \\
&=-m\ddot z_0(t)\la \tilde \Psi_{\rm I}(t)|\hat J_z(t) 
|\tilde \Psi_{\rm I}(t)\ra.
\end{align}
Substituting Eq.~(\ref{tPsi}) into this result, one finds
\beq
\frac{d\tilde E}{dt}=-m^2\ddot z_0(t)\int_{-\infty}^\infty dt' 
\Pi_{zz}(t-t') \dot z_0(t'),
\label{dtildeE_dt}
\eeq
which differs from the correct result in Eq.~(\ref{edrpmsys}).
For the sinusoidal displacement, we have
\beq
\frac{d\tilde E}{dt}
=m^2z_0^2\omega^3\sin\omega t \left [\cos\omega t \,
{\rm Re}\Pi_{zz}(\omega)+\sin\omega t \,{\rm Im}
\Pi_{zz}(\omega) \right ],
\label{eartokuno}
\eeq
which clearly has a different time dependence from $dE/dt$ in
Eq.~(\ref{edrpmsin}).
However, when averaged over one period, the energy 
absorption rate is 
\beq
\overline{\frac{d \tilde E}{dt}}=\frac{1}{2}m^2z_0^2\omega^3
{\rm Im}\Pi_{zz}(\omega),
\eeq
which is the same as $\overline{dE/dt}$. Thus the approach
adopted in Ref.~\cite{Tokuno} does indeed yield the correct 
time-averaged energy absorption rate for a sinusoidal displacement.
However, this is not true for other forms of the
displacement. For example, for the displacement
$z_0(t)=\theta(t) v_0 t $, which corresponds to the external
potential moving with a constant velocity $v_0$ for $t>0$,
Eq.~(\ref{edrtokuno}) gives 
\begin{align}
\frac{d\tilde E}{dt}&=\la\tilde\Psi(t)|(-m{\ddot z}_0(t)
\hat J_z)|\tilde\Psi(t)\ra \nn \\
&=-mv_0\delta (t) \la \tilde\Psi(0)|\hat J_z|\tilde \Psi(0)\ra \nn \\
&=0
\end{align}
since the initial state is one in which there is no current. 
This conclusion also follows from
Eq.~(\ref{dtildeE_dt}). On the other hand, Eq.~(\ref{edrpmsys})
gives
\begin{align}
\frac{dE}{dt}=m^2v_0^2\frac{d}{dt}\int_0^t
dt'\Pi_{zz}(t-t') = m^2v_0^2 \Pi_{zz}(t),
\end{align}
which is a non-zero result.
This shows that Eq.~(\ref{dtildeE_dt}) cannot be the 
correct result for the energy absorption rate in general.
\section{}
In this appendix we provide an alternative explanation for 
why the energy
absorption rate in Eq.~(\ref{ear}) vanishes when $\omega = 
\omega_z$. To this end, we write the Hamiltonian in
Eq.~(\ref{H_shaking}) as
\beq
\hat H_{\rm ad}(t) = \hat H + e^{\eta t} \hat H'(t,\lambda)
\eeq
where
\beq
\hat H'(t,\lambda) = \sum_{i=1}^N \left [ V_{\rm ext}(\hat \br_i 
- \lambda \bx(t)) - V_{\rm ext}(\hat \br_i) \right ].
\eeq
The perturbation $\hat H'(t,\lambda)$ is turned on adiabatically via 
the parameter $\eta$. For $t \to -\infty$, $\hat H_{\rm ad}(t)$ 
reduces to $\hat H$, while for $\eta \to 0$ one recovers the 
Hamiltonian $\hat H(t)$ in Eq.~(\ref{H_shaking}). 
The parameter $\lambda$ is introduced
as an ordering parameter in the perturbation analysis of
$\hat H'(t,\lambda)$ and is set to unity at the end of the calculation. 
The displacement will be taken to have the specific form
$\bx(t) = \hat {\bf z} z_0 \sin \omega_z t$. We observe that 
$\hat H'(t,\lambda)$ can be written as
\begin{align}
\hat H'(t,\lambda)=\hat T(\lambda \bx(t),\lambda\bp(t))\hat V_{\rm 
ext}\hat T^\dag(\lambda\bx(t),\lambda\bp(t))-\hat V_{\rm ext},
\end{align}
where $\hat V_{\rm ext}=\sum_{i=1}^N V_{\rm ext}(\hat \br_i)$.
Although the required displacement of $\hat V_{\rm ext}$ can be 
generated with any $\bp(t)$, we make the choice $\bp(t)=Md\bx(t)/dt$ 
to ensure that
the displacement operator evolves in time according to
Eq.~(\ref{emt}).
This property will be shown to be crucial in the derivation of
our final result.

We now construct the dynamical state in the interaction picture
which reduces to $|\Phi_0\ra$, the ground state of $\hat H$, in
the $t \to -\infty$ limit. This state is given by 
\begin{align}
|\Psi_{\rm I}(t)\ra=\hat\calU_{\rm I}(t,\lambda)|\Phi_0\ra,
\end{align}
where the evolution operator satisfies the integral equation
\begin{align}
\hat\calU_{\rm I}(t,\lambda)
=1 +\frac{1}{i\hbar}\int_{-\infty}^t dt_1 e^{\eta t_1}
\hat H_{\rm I}'(t_1,\lambda)\hat\calU_{\rm I}(t_1,\lambda).
\label{evolop}
\end{align}
Here  
\begin{align}
\hat H_{\rm I}'(t,\lambda)&= e^{i\hat H t/\hbar}\hat 
H'(t,\lambda)e^{-i\hat H t/\hbar} \nn \\
&=\hat T_{\rm I}(\lambda\bx(t),\lambda\bp(t))\hat V_{\rm ext,\rm 
I}(t)\hat T_{\rm I}^\dag(\lambda\bx(t),\lambda\bp(t))-
\hat V_{\rm ext, \rm I}(t),
\label{HPMlam}
\end{align}
where $\hat V_{\rm ext, \rm I}(t) = e^{i\hat H t/\hbar}
\hat V_{\rm ext} e^{-i\hat H t/\hbar}$ and
\begin{align} 
 \hat T_{\rm I}(\lambda\bx(t),\lambda\bp(t))&\equiv e^{i\hat H 
t/\hbar}\hat T(\lambda\bx(t),\lambda\bp(t))e^{-i\hat H t/\hbar} \nn \\
&=\exp\left\{ \frac{i}{\hbar}\lambda\left (\bp(t)\cdot\hat \bR_{\rm 
I}(t)-\bx(t)\cdot\hat \bP_{\rm I}(t) \right )\right \} \nn \\
&\equiv \exp\left \{\frac{i}{\hbar}\lambda \hat A_{\rm I}(t)\right\},
\end{align}
with
\beq
\hat A (t)=\bp(t)\cdot\hat \bR-\bx(t)\cdot\hat \bP.
\label{Aop}
\eeq

We now evaluate $\hat\calU_{\rm I}(t,\lambda)$ explicitly. Expanding 
this operator in powers of $\lambda$, we have
\beq
\hat\calU_{\rm I}(t,\lambda)=\sum_{n=0}^\infty 
\frac{1}{n!}\hat\calU^{(n)}_{\rm I}(t,0)\lambda^n,
\label{Uexp}
\eeq
where
\beq
\hat\calU^{(n)}_{\rm I}(t,0)\equiv \left .\frac{\pa^n}{\pa 
\lambda^n}\hat\calU_{\rm I}(t,\lambda)\right |_{\lambda =0}.
\eeq
Similarly we have
\beq
\hat H_{\rm I}'(t,\lambda)= \sum_{n=0}^\infty \frac{1}{n!}\hat 
H_{\rm I}'^{(n)}(t,0) \lambda^n,
\label{Hpexp}
\eeq
where 
\beq
\hat H_{\rm I}'^{(n)}(t,0)\equiv \left .\frac{\pa^n}{\pa \lambda^n}\hat 
H_{\rm I}'(t,\lambda)\right |_{\lambda =0}.
\label{HIn}
\eeq
Using Eq.~(\ref{HPMlam}) in Eq.~(\ref{HIn}), we obtain
\beq 
\hat H_{\rm I}'^{(0)}(t,0)=0,
\eeq
\begin{align}
\hat H_{\rm I}'^{(1)}(t,0)&=\frac{i}{\hbar}[\hat A_{\rm I}(t),\hat 
V_{\rm ext,\rm I}(t)],
\label{HPM1}
\end{align}
 and for $n>1$,
\beq
\hat H_{\rm I}'^{(n)}(t,0)=\frac{i}{\hbar}[\hat A_{\rm I}(t),\hat 
H_{\rm I}'^{(n-1)}(t,0)].
\eeq

Substituting Eqs.~(\ref{Uexp}) and (\ref{Hpexp}) into 
Eq.~(\ref{evolop}) and comparing like powers of $\lambda$, we find
\beq
\hat\calU^{(n)}_{\rm 
I}(t,0)=\frac{1}{i\hbar}\sum_{m=0}^n\frac{n!}{m!(n-m)!}\int_{-\infty}^t 
dt_1 e^{\eta t_1}\hat H_{\rm I}'^{(m)}(t_1,0)
\hat\calU^{(n-m)}_{\rm I}(t_1,0).
\label{calUn}
\eeq
Let us first consider $\hat\calU^{(1)}_{\rm I}(t,0)$. From 
Eq.~(\ref{HPM1}) we have
\begin{align}
\hat H_{\rm I}'^{(1)}(t,0)
&=-\frac{imz_0(t)}{\hbar}e^{i\hat H t/\hbar}[\hat J_{z}(t),
\hat V_{\rm ext}(t)] e^{-i\hat H t/\hbar}\nn \\
&=-\frac{imz_0(t)}{\hbar}e^{i\hat H t/\hbar}\left [\hat J_{z}, \hat 
H-\sum_{i=1}^N  \frac{\hat \bp_i^2}{2m}-\sum_{i<j}^Nv(\hat \br_i-\hat 
\br_j)-\sum_{i=1}^N V_{\rm tr}(\hat \br_i)\right]e^{-i\hat H 
t/\hbar}\nn \\
&= mz_0(t)\left \{ \frac{1}{i\hbar}[\hat J_{z,\rm I}(t), \hat H]+ 
N\omega_z^2\hat R_{z,\rm I}(t)\right \} \nn \\
&=Nmz_0(t)\left \{ \frac{\pa^2 }{\pa t^2}\hat R_{z,\rm I}(t)+ 
\omega_z^2\hat R_{z,\rm I}(t)\right \}.
\label{HIp1}
\end{align}
Using Eq.~(\ref{Aop}) and $z_0(t)=z_0\sin\omega_z t$, we find
\beq
\hat H_{\rm I}'^{(1)}(t,0)=-\frac{\pa }{\pa t}\hat A_{\rm I}(t).
\label{HPM1f}
\eeq
This result is {\it only} true when $\bp(t) = Md\bx(t)/dt$, as
assumed.
Inserting Eq.~(\ref{HPM1f}) into Eq.~(\ref{calUn}) for $n=1$, we have
\begin{align}
 \hat\calU^{(1)}_{\rm I}(t,0)&=\frac{1}{i\hbar}\int_{-\infty}^t dt_1 
e^{\eta t_1}\hat H_{\rm I}'^{(1)}(0,t_1) \nn \\
&=\frac{i}{\hbar} \hat A_{\rm I}(t) + {\cal O}(\eta),
\label{calU1}
\end{align}
where ${\cal O}(\eta)$ denotes terms that vanish in the $\eta
\to 0$ limit.

Repeating this calculation for $n=2$, we find
\begin{align}
\hat\calU^{(2)}_{\rm I}(t,0)&=\frac{1}{i\hbar}\int_{-\infty}^t dt_1 
e^{\eta t_1} \left \{\hat H_{\rm I}'^{(2)}(t_1,0) \hat\calU^{(0)}_{\rm 
I}(t_1,0)+ 2\hat H_{\rm I}'^{(1)}(t_1,0) \hat\calU^{(1)}_{\rm
I}(t_1,0) \right \}\nn \\
&=-\frac{1}{\hbar^2}\int_{-\infty}^t dt_1 e^{\eta t_1} \left \{[\hat A_{\rm 
I}(t_1),\frac{\pa }{\pa t}\hat A_{\rm 
I}(t_1)]+2 \frac{\pa \hat A_{\rm I}(t_1)}{\pa 
t} \hat A_{\rm I}(t_1)\right \} \nn \\
&=-\frac{1}{\hbar^2}\int_{-\infty}^t dt_1 e^{\eta t_1}\frac{\pa 
\hat A ^2_{\rm I}(t_1) }{\pa t_1} \nn \\
&=\left (\frac{i}{\hbar}\hat A_{\rm I}(t)\right )^2 + {\cal
O}(\eta)
\end{align}
The results for $n=1$ and $n=2$ suggest that
\beq
\hat\calU^{(n)}_{\rm I}(0,t)=\left(\frac{i}{\hbar}\hat 
A_{\rm I}(t) \right )^n+ {\cal O}(\eta)
\label{calUna}
\eeq
for all $n\geq 1$. This in fact can be proven by induction.
Without presenting the details, we thus find that in the $\eta \to 0$
limit,
\begin{align}
\hat\calU_{\rm I}(t,\lambda)&=\exp\left\{\frac{i}{\hbar}\hat A_{\rm 
I}(t)\right\} \nn \\
&=e^{i\hat Ht/\hbar}\hat T(\lambda\bx(t),\lambda\bp(t))e^{-i\hat 
Ht/\hbar}.
\end{align}
The dynamic state of interest is thus given by
\begin{align}
|\Psi(t)\ra&=e^{-i\hat Ht/\hbar}|\Psi_{\rm I}(t)\ra\nn \\
&=e^{-i\hat Ht/\hbar}\hat \calU_{\rm I}(t,\lambda)|\Phi_0\ra \nn \\
&=e^{-i\hat E_0t/\hbar}\hat T(\lambda\bx(t),\lambda\bp(t))|\Phi_0\ra.
\end{align}
We have thus proved that, after the perturbation is switched on 
adiabatically, the final state is the ground state of $\hat H$
oscillating together with the external potential.

The energy of the system in the state $|\Psi(t)\ra$ is 
\beq
E(t,\lambda)=\la\Psi(t)|\hat H_0+\sum_{i=1}^NV_{\rm
ext}(\hat\br_i-\lambda \hat \bz z_0(t))|\Psi(t)\ra.
\eeq
This energy can be obtained from Eq.~(\ref{tildeE}) with the
transcription $\bx_0(t) \to -\lambda z_0(t)\hat \bz$, giving
\beq
E(t,\lambda )=E_0 +\lambda^2 E_{\rm
cm}+\lambda \la\Phi_0|\hat P_z|\Phi_0\ra \dot z_0(t) +
\lambda M\omega_z^2\la\Phi_0|\hat R_z|\Phi_0\ra z_0(t).
\label{E_lambda}
\eeq
The terms in Eq.~(\ref{E_lambda})
linear in $\lambda$ oscillate harmonically at the frequency
$\omega_z$ and do not contribute to the time-averaged energy
absorption rate. We thus find 
\beq
\overline{\frac{\partial E(t,\lambda)}{\partial t}} = 0.
\eeq
This result is true to all orders in $\lambda$ and in
particular, demonstrates that the linear response energy
absorption rate (of order $\lambda^2$) vanishes when $\omega =
\omega_z$.

\end{document}